\documentclass[a4paper,11pt]{article}
\pdfoutput=1 

\usepackage{jcappub} 

\usepackage[T1]{fontenc} 
\usepackage[utf8]{inputenc}

\usepackage{graphicx}
\usepackage{amsmath}
\usepackage{amssymb}
\usepackage{color}
\usepackage{cancel}
\usepackage{hyperref}

\newcommand{\dd}{\mathrm{d}}
\newcommand{\cH}{\mathcal{H}}
\newcommand{\xlat}[3]{\mathbf{x}_{\mathbf{i#1},\mathbf{j#2},\mathbf{k#3}}}
\newcommand{\taulat}[1]{\tau_\mathbf{n#1}}
\newcommand{\lat}[5]{#1^{\mathbf{#2}}_{\mathbf{i#3},\mathbf{j#4},\mathbf{k#5}}}
\newcommand{\latFT}[2]{\tilde{#1}^{\mathbf{#2}}_{\mathbf{u},\mathbf{v},\mathbf{w}}}
\newcommand{\klat}[1]{k^{#1}_{\mathbf{u},\mathbf{v},\mathbf{w}}}

\newcommand{\lf}{{\sc lat}field{\sc 2}}

\title{\boldmath gevolution: a cosmological N-body code based on General Relativity}

\author[a,b,1]{Julian Adamek,\note{Corresponding author.}}
\author[c,b]{David Daverio,}
\author[b]{Ruth Durrer}
\author[b]{and Martin~Kunz}

\affiliation[a]{Laboratoire Univers et Th\'eories -- UMR 8102, Observatoire de Paris, 5 Place Jules Janssen, 92195 Meudon CEDEX, France}
\affiliation[b]{D\'epartement de Physique Th\'eorique \& Center for Astroparticle Physics, Universit\'e de Gen\`eve, 24 Quai E.\ Ansermet, 1211 Gen\`eve 4, Switzerland}
\affiliation[c]{African Institute for Mathematical Sciences, 6 Melrose Road, Muizenberg 7945, South Africa}

\emailAdd{julian.adamek@obspm.fr}
\emailAdd{david.daverio@unige.ch}
\emailAdd{ruth.durrer@unige.ch}
\emailAdd{martin.kunz@unige.ch}

\abstract{We present a new N-body code, \textit{gevolution}, for the evolution of large scale structure in the Universe.
Our code is based on a weak field expansion of General Relativity and calculates all six metric degrees of freedom in
Poisson gauge. N-body particles are evolved by solving the geodesic equation which we write in terms of a canonical momentum
such that it remains valid also for relativistic particles. We validate the code by considering the Schwarzschild solution and,
in the Newtonian limit, by comparing with the Newtonian N-body codes \textit{Gadget-2} and \textit{RAMSES}. We then proceed with a simulation of
large scale structure in a Universe with massive neutrinos where we study the gravitational slip induced by the neutrino shear stress. 
The code can be extended to include different kinds of dark energy or modified gravity models and going beyond the usually adopted quasi-static approximation. Our code is publicly available.

\bigskip

{\footnotesize This is an author-created, un-copyedited version of an article published in JCAP. IOP Publishing Ltd is not responsible for any errors or omissions in this version of the manuscript or any version derived from it. The Version of Record is available online at \texttt{\url{http://dx.doi.org/10.1088/1475-7516/2016/07/053}}.}}

\begin{document}
\maketitle
\flushbottom

\section{Introduction}
\label{sec:intro}

Almost to the day 100 years ago, Einstein presented his theory of General Relativity (GR) to the Royal Prussian Academy of Sciences \cite{Einstein:1915ca}.
This completely transformed our view on gravitation. Spacetime is no longer understood as an absolute element, a fixed stage on which interactions take place,
but as a dynamical entity which takes part in the interactions. Nowhere is this more apparent than in cosmology, where the dynamics imply that our Universe
can not be in an eternal ``steady-state''. From Lema\^\i tre's and Hubble's discovery of the expansion law \cite{Lemaitre:1927zz,Hubble:1929ig} to the latest precision measurements of cosmic
microwave background anisotropies \cite{Adam:2015rua} we have collected ample evidence that our Universe has expanded and evolved from a very hot
and fairly uniform initial state (the hot big bang which probably was preceded by an inflationary phase) into the present cold and highly differentiated state.
The fact that all current observations can be consistently interpreted in the context of GR, giving rise to the $\Lambda$CDM concordance model of cosmology, is arguably
one of the greatest successes of Einstein's theory of gravity.

This concordance model achieves a good fit to the data only by two phenomenological postulates, the existence of two ``dark'' contributors of stress-energy.
Non-baryonic Dark Matter (DM) outweighs ordinary matter (atoms and ionized gas) by more than 5:1 and is therefore the dominant driver for gravitational clustering.
It is usually modelled as a yet-to-be-found weakly interacting massive particle. Dark Energy (DE) is an even more exotic entity which effectively acts like
a cosmological constant and gives rise to accelerated expansion. One should keep in mind that evidence for these dark components
remains circumstantial at this point, as we only observe their apparent gravitational effects. Major efforts are being undertaken to obtain a deeper understanding
of DM and DE at the fundamental level. In particular the study of cosmic large scale structure with high precision seems to be a promising avenue, and many
upcoming surveys \cite{Carilli:2004nx,Abell:2009aa,Laureijs:2011gra} are targeting more precise measurement of DM and DE properties via their gravitational interaction.
Other experiments, so called \textit{direct} and \textit{indirect} DM searches, aim to detect DM also via non-gravitational interactions~\cite{Baudis:2015mpa}.

The theoretical modelling of the formation of large scale structure is very challenging at nonlinear scales. Computer simulations have a long tradition in this field, and due to
technological advancement, continue to keep pace with the increasing data quality of astronomical observations. However, these simulations generally still use
Newton's law of gravitation. This has spurred an extensive debate in the literature whether and how this approximation can be justified in the current era of
precision cosmology \cite{Chisari:2011iq,Green:2011wc,Flender:2012nq,Rigopoulos:2013nda,Fidler:2015npa}. It seems that the Newtonian approach works fairly well
within the context of the $\Lambda$CDM model. Firstly, if one neglects radiation in the late Universe, the only source of perturbations is nonrelativistic matter.
A Newtonian scheme would clearly be unsuitable for relativistic sources. Furthermore, there is a separation of scales between the nonlinear scale and the
cosmological horizon. A Newtonian scheme is ignorant about the presence of a horizon, but it turns out that linear perturbations of nonrelativistic matter
can be identified with quantities of linear perturbation theory of GR. One can therefore find a consistent interpretation of Newtonian simulations also at very large
scales \cite{Fidler:2015npa}, but only at the linear level and one has to be very careful about gauge issues. The presence of radiation also has a small effect on the evolution of
structure which is difficult to account for within a Newtonian framework.

Despite these subtleties, models beyond $\Lambda$CDM have been studied with simulation codes using some modified Newtonian approaches
\cite{Schmidt:2009sg,Li:2011vk,Puchwein:2013lza,Llinares:2013jza}. These codes are customized towards specific models where appropriate approximations (such as the quasi-static limit which is only valid on small scales and for small enough speed of sound, see
\cite{Sawicki:2015zya} for a discussion) can be justified. However, a general framework would be desirable in order to access the entire space of viable models. We have therefore developed a new
N-body code which is completely based on GR \cite{Adamek:2015eda}. The aim of this paper is to introduce the code and explain in detail its theoretical underpinning. As we make our code
available to the community, this paper also serves as a first introduction for future users. From the very beginning we  avoid any reference to the Newtonian picture
and follow a conceptually clean approach which ensures self-consistency and compliance with GR principles at every step. The central element of our framework is
a weak field expansion, meaning that we are able to treat any settings where no strong gravitational fields appear. This, of course, includes any setting where
the Newtonian approximation would be applicable, but also arbitrary scenarios with relativistic sources as long as gravitational fields are not very strong. The framework is well suited for cosmology, but it could also be fruitful for astrophysical applications with
moderate gravitational fields where a Newtonian treatment is insufficient.

Let us also clarify the relation between our framework and other numerical approaches such as the ones of \cite{Bentivegna:2015flc,Mertens:2015ttp}. While
those approaches do not rely on a weak field expansion, allowing access to the strong field regime, they are based on a fluid description
of the matter sources. The fluid approximation is valid for DM only as long as nonlinear collapse has not yet caused the
trajectories of mass elements to cross each other, at which point the fluid equations become singular. This phenomenon, sometimes called shell-crossing, is ubiquitous during the process of nonlinear
clustering of effectively collisionless matter. It is an essential aspect of hierarchical structure formation and the virialization of
DM halos. In practice this means that approaches based on the fluid description are expected to break down at the nonlinear scale
while a weak field N-body scheme should be able to cover all the scales down to the occurrence of strong fields. In the present Universe
the nonlinear scale is at some tens of megaparsecs while the largest strong field regions\footnote{Our notion of strong fields is actually
gauge dependent. We are working in Poisson gauge where metric perturbations become large as one approaches the Schwarzschild
radius. However, the tidal forces are very small at the Schwarzschild horizon of a supermassive black hole, and one could find
a gauge where this region would be classified as weak field regime.} are supermassive black holes with a radius of a
few hundred astronomical units, i.e.\ in the range of milliparsecs. Therefore, while certainly interesting for the study of some idealized setups, the fluid approaches are of limited use for the modelling of realistic cosmological structures.

In Section \ref{sec:EE} we derive the equations which govern the evolution of the metric and hence of spacetime geometry. These are formulated in a
way which is convenient for numerical integration but also intuitive for any cosmologist familiar with relativistic perturbation theory. Section \ref{sec:particles}
discusses the ensemble of N-body particles, in particular its stress-energy and its evolution under gravity. The numerical implementation is outlined
in Section \ref{sec:code}. Some first simulation results are presented in Section \ref{sec:simulations}, ranging from verification checks to cosmological applications.
We carry out some simulations with relativistic particle species, a setting for which Newtonian codes are severely limited. Our results are summarized in
Section \ref{sec:summary}. Some technical aspects of our code are explained in three appendices. More detailed instructions for users of the code can be found in the public release at
\texttt{\url{https://github.com/gevolution-code/gevolution-1.0.git}}.

\section{Einstein's equations}
\label{sec:EE}

We start by writing an ansatz for the line element, which for our purposes shall take the form of a perturbed Friedman-Lema\^itre-Robertson-Walker (FLRW)
metric in Poisson gauge,
\begin{equation}
\label{eq:metric}
 \dd s^2 = g_{\mu\nu} \dd x^\mu \dd x^\nu = a^2(\tau) \left[-\left(1+2\Psi\right) \dd\tau^2 - 2 B_i \dd x^i \dd\tau + \left(1-2\Phi\right) \delta_{ij} \dd x^i \dd x^j + h_{ij} \dd x^i \dd x^j\right] \, ,
\end{equation}
where $a$ is a background scale factor (the choice of background is discussed below), $\tau$ is conformal time, and $x^i$ are comoving Cartesian coordinates.
As usual we  follow the convention that summation is implied for repeated indices, where letters from the Latin alphabet specify space-like indices and letters
from the Greek alphabet run over all four spacetime dimensions. We also use the shorthands $f' \doteq \partial f / \partial \tau$ and
$f_{,i} \doteq \partial f / \partial x^i$ to denote partial derivatives, $\Delta f \doteq \delta^{ij} f_{,ij}$ to denote the spatial Laplace operator, and
$f_{(ij)} \doteq (f_{ij} + f_{ji}) / 2$ to denote index symmetrization.

The Poisson gauge is usually introduced at linear level, but we  maintain the gauge conditions $\delta^{ij} B_{i,j} = \delta^{ij} h_{ij} = \delta^{jk} h_{ij,k} = 0$
even though we go beyond perturbation theory. Our gauge conditions can be maintained as long as we remain in our weak-field setting where all gravitational fields ($\Psi$, $\Phi$,
$B_i$ and $h_{ij}$) are small. In fact, our expansion remains linear in $B_i$ and $h_{ij}$ (though not in $\Psi$, $\Phi$) such that they can be directly interpreted
as the usual spin-1 and spin-2 perturbations familiar from perturbation theory. In other words, $B_i$ is the spin-1 field responsible for frame dragging, whereas
$h_{ij}$ carries the two spin-2 degrees of freedom of gravitational waves or more general tensor perturbations.

The metric of eq.~(\ref{eq:metric}) is of course motivated having a cosmological application in mind. However, if one wants to use our framework to describe perturbations
around a Minkowski geometry instead of FLRW, as would be appropriate for astrophysical applications, one can simply set $a =1$ and $\cH \doteq a'/a = 0$
in all our equations. An example is discussed in Section \ref{sec:pointmass}.

The metric variables, i.e.\ $a(\tau)$ for the background and $\Psi$, $\Phi$, $B_i$, $h_{ij}$ for the perturbations, are evolved according to Einstein's field equations,
\begin{equation}
\label{eq:EFE}
 G^\mu_\nu = 8 \pi G T^\mu_\nu \, ,
\end{equation}
where $G^\mu_\nu$ is the Einstein tensor (a quantity built from the metric and its first and second derivatives) and $T^\mu_\nu$ is the stress-energy tensor. For a generic model
we write the total action as
\begin{equation}
 \mathcal{S} = \mathcal{S}_{\mathrm{EH}} + \mathcal{S}_\mathrm{m} = \frac{1}{16 \pi G} \int\! \sqrt{-g} \mathcal{R} \dd^4 x + \int\! \sqrt{-g} \mathcal{L}_\mathrm{m} \dd^4 x \, ,
\end{equation}
where $g$ is the metric determinant, $\mathcal{R}$ is the Ricci scalar and $\mathcal{L}_\mathrm{m}$ is the matter Lagrangian which usually depends on $g_{\mu\nu}$ and matter variables.
We use the convention that a possible cosmological constant would be included in the matter Lagrangian. The action principle requires that the variation with respect to $\delta g_{\mu\nu}$ vanishes,
which implies
\begin{equation}
 -\frac{\delta\mathcal{R}}{\delta g_{\mu\nu}} - \frac{\mathcal{R}}{\sqrt{-g}}\frac{\delta\sqrt{-g}}{\delta g_{\mu\nu}} = \frac{16 \pi G}{\sqrt{-g}}\frac{\delta\left(\sqrt{-g}\mathcal{L}_\mathrm{m}\right)}{\delta g_{\mu\nu}} \, .
\end{equation}
Working out the left hand side, we obtain $G^{\mu\nu} \doteq \mathcal{R}^{\mu\nu} + \frac{1}{2} g^{\mu\nu} \mathcal{R}$, where $\mathcal{R}^{\mu\nu}$ is the Ricci tensor.
This means that the stress-energy tensor is related to the matter Lagrangian by
\begin{equation}
\label{eq:genericTmunu}
 T^{\mu\nu} \doteq \frac{2}{\sqrt{-g}}\frac{\delta\left(\sqrt{-g}\mathcal{L}_\mathrm{m}\right)}{\delta g_{\mu\nu}} \, .
\end{equation}
In the following subsections we explain how eq.~(\ref{eq:EFE}) can be written in a convenient form by using our ansatz for the metric, eq.~(\ref{eq:metric}), and employing
a judicious expansion in the metric perturbation variables.

\subsection{Choice of background}
\label{sec:background}

In linear perturbation theory the background is usually defined as the model which is obtained in the limit where perturbations are taken to zero. This is not necessarily
a good prescription beyond the linear level, since there are nonlinear terms which do not average to zero on the space-like hypersurface and therefore could give a
relevant correction to the ``average'' evolution of the geometry. This is one aspect of the well known ``backreaction'' problem (see \cite{Buchert:2011sx,Clarkson:2011zq}
for recent reviews and \cite{Adamek:2014gva} for a discussion in the present framework). Instead of trying to address this problem by giving a prescription of how to
construct the background in general, we  follow a different philosophy. We note that there is a residual gauge freedom which allows for some degree of arbitrariness
in the background model: a small change to the background $a(\tau) \rightarrow \tilde{a}(\tau) = a(\tau) + \delta a(\tau)$ can be absorbed into the perturbation
variables to leave the line element~(\ref{eq:metric}) invariant. Evidently, $\Psi$ and $\Phi$ may acquire a homogeneous mode by this gauge transformation, but we 
allow for this as long as this homogeneous mode remains a small perturbation and does not spoil the weak field expansion which is elaborated in the next subsection.

Effectively we conjecture that, even though perturbations of the \textit{stress-energy} can become arbitrarily large, there may still exist a coordinate
system and a choice of background model such that the geometry can be described by eq.~(\ref{eq:metric}) where all \textit{metric} perturbation variables are small. Since we start our
simulations at a time where perturbation theory is still valid, we know that this conjecture holds
initially\footnote{There are cases where linear perturbation theory is valid (and the geometry therefore remains close to FLRW)
but $\Phi$ and $\Psi$ become large. But this can happen only on super-Hubble scales as $C/R \simeq k^2(\Phi+\Psi)/\cH^2$ must remain small
for perturbation theory to be valid. Here $C$ denotes a typical component of the Weyl curvature while $R$ is a typical component of the background Ricci tensor. This, however, means that it is a gauge other than the Poisson gauge in which the perturbation variables remain
small. An example of this behavior is dilaton inflation~\cite{Brustein:1994kn}. Such scenarios are not the topic of this work.}. We can monitor the perturbations during the evolution to
make sure that the weak field expansion remains valid at all times. If, for instance, our background model turns out to be inadequate, the system reacts by generating
large homogeneous modes in $\Phi$ and $\Psi$ (see Section \ref{sec:backreaction} for an explicit demonstration) which eventually would break the scheme. In this case the background model has to be improved. We do not provide a general set of conditions under which
an appropriate background can be found, but we can address the backreaction problem on a case-by-case basis by solving the equations and monitoring
the size of the geometric perturbation variables. On the other hand, the smallness of $\Phi$ and $\Psi$ (and of the other metric perturbations) is a sufficient condition to ensure that
the geometry is close to the chosen background and that the weak field expansion discussed in the next section is valid.

Let us now turn to the aforementioned residual gauge freedom.
We can actually exploit this gauge freedom to make a convenient choice for the background model. We only have to make sure that the homogeneous modes
in the perturbations remain small enough such that we still trust our evolution equations. The background model (and associated parameters such as $\cH$) therefore is in this sense
gauge dependent, but of course observables are not. Once an experimental setup is specified, the outcome is independent of the choice of gauge. This choice just specified which part is considered as belonging to the background and which part as belonging to the perturbations. We make a convenient choice for the background stress-energy tensor
$\bar{T}^\mu_\nu$, which determines the scale factor according to Friedmann's equation,
\begin{equation}
 \label{eq:Friedmann}
 -3 \frac{\cH^2}{a^2} = 8 \pi G \bar{T}^0_0 \, .
\end{equation}
The perturbations are then solved by using Einstein's equations with the background contributions subtracted, for instance
\begin{equation}
\label{eq:E00}
 G^0_0 + 3 \frac{\cH^2}{a^2} = 8 \pi G \left(T^0_0 - \bar{T}^0_0\right) \, ,
\end{equation}
which does not introduce any assumptions about the background model. We only require that the metric perturbations $\Phi$, $\Psi$, $B_i$ and $h_{ij}$ remain small
on this background. The homogeneous modes of the perturbation variables are computed consistently
such that the observables do not depend on the precise choice of $\bar{T}^\mu_\nu$. A numerical study of this issue is presented in Section \ref{sec:backreaction}.

We finally note that we still have not exhausted the gauge freedom completely. We can rescale our spatial coordinates such that the homogeneous mode
of $\Phi$ takes a particular value at a given  instance in time. This can be done only once, the homogeneous mode at all other times being determined
by the dynamics. We  use this freedom to set the homogeneous mode of $\Phi$ to zero at the initial time of the simulation. We can also rescale the time
coordinate to redefine the homogeneous mode of $\Psi$. As opposed to the previous case the dynamics do not determine the evolution of this mode. Instead,
we can fix any functional form of this mode, and the dynamics of all other variables is then automatically solved with respect to the corresponding
choice of time coordinate. For convenience, in our code the homogeneous mode of $\Psi$ is set equal to the one of $\Phi$, so that the homogeneous mode of $\Phi-\Psi$ vanishes.
Again, these choices do not have any effect at the level of observables.

\subsection{Weak field expansion}
\label{sec:weak-field}

In order to reduce the ten nonlinear coupled partial differential equations of eq.~(\ref{eq:EFE}) to a tractable set of evolution equations for the perturbation variables,
we  employ a weak field expansion. The assumption behind this scheme is that there are no strong gravitational fields at the scales of interest. It should be stressed
that this does not imply that the perturbations of the stress-energy tensor, which are the sources of the gravitational fields, have to remain small. In fact, $\vert T^0_0 \vert$
is generally  much larger than $\vert \bar{T}^0_0 \vert$ inside dense regions. As an example, our solar system perfectly fits into a weak field description, despite the fact that
the density varies by some 25 orders of magnitude between the center of the sun ($\simeq 162$ g/cm$^3$) and the average matter density in the interplanetary medium ($\simeq 10^{-23}$ g/cm$^3$ at $1$ AU from the sun). A weak field expansion amounts to an expansion
in terms of the metric perturbations alone, without at the same time employing any expansion of the matter variables. It is not equivalent to a post-Newtonian expansion,
which is an expansion in inverse powers of the speed of light. Such an expansion would be suitable only as long as the perturbations of stress-energy are non-relativistic to
a good approximation. We do not place this restriction on our stress-energy tensor (the relation between our scheme and a post-Newtonian one \cite{Milillo:2015cva} is discussed in more detail in
\cite{Adamek:2014xba}).

Empirically we know that the scalar perturbations $\Psi$, $\Phi$ are generally much larger than the spin-1 and spin-2 perturbations $B_i$ and $h_{ij}$. This is why
Newtonian simulations could enjoy such a successful history. While the gravitational potential can be easily observed with a table-top experiment \cite{Chou:2010zz}, there are only
few experiments which were able to detect frame dragging directly (e.g.\ \cite{Ciufolini:2004rq,Everitt:2011hp}), and a direct observation of gravitational waves has only recently been achieved \cite{Abbott:2016blz} as the result of a remarkable technological feat.
When expanding Einstein's equations~(\ref{eq:EFE}) in terms of the metric perturbations, we therefore keep only the linear terms\footnote{In our previous works
\cite{Adamek:2014xba,Adamek:2013wja} we claimed that we keep all quadratic terms with two spatial derivatives, which would have included terms built with $B_i$ or
$h_{ij}$. In fact, we then dropped these terms from the equations without mention. A technical justification for this step is given in \cite{Green:2011wc}, but it relies
on some restrictive assumptions about the stress-energy tensor which we would like to relax. We admit that a general model could have relativistic sources creating large
spin-1 and spin-2 metric perturbations, but when quadratic terms in these variables become relevant, we do not consider this a weak field setting anymore and therefore
our framework would be inappropriate.} for $B_i$ and $h_{ij}$. However, for the scalar potentials $\Psi$ and $\Phi$ we are more cautious.
We assume that the
potentials are small everywhere, but we admit that they have fluctuations at small scales which can lead to large density fluctuations: $\delta\rho/\rho \sim (k/\cH)^2\Phi$ can become much larger than unity. This reflects the Newtonian gravitational instability, which is also the only instability of General Relativity. Curvature, which is related to the spatial Laplacian of $\Phi$ and $\Psi$
therefore becomes large at small scales. In order to appreciate this effect, we  keep
the quadratic terms of $\Psi$ and $\Phi$ with the highest number of spatial derivatives. Since the differential equations are second order, the highest possible number is two.

However, as we will see below, the \textit{difference} of the two potentials, 
\begin{equation} \chi \doteq \Phi-\Psi\,,\end{equation} 
is determined from the same set of equations as $B_i$ and $h_{ij}$.
We therefore treat $\chi$ on the same footing as the spin-1 and spin-2 perturbations. In other words, while we keep some quadratic terms in $\Phi$, after systematically replacing $\Psi$ by $\Psi=\Phi-\chi$, we will
only keep terms linear in $\chi$.

To summarize, we keep all terms up to linear order in the metric perturbations without distinction, but from the quadratic ones we only keep the ones built with
$\Phi$ which have exactly two spatial derivatives. Examples are $\Phi\Delta\Phi$ or $\delta^{ij} \Phi_{,i} \Phi_{,j}$. Terms like $\Phi \Phi'$, $\delta^{ij}
B_i \Delta B_j$ or all other higher-order terms are subleading and are dropped. We argue that this truncation contains all the relevant terms to compute $\chi$, $B_i$ and
$h_{ij}$ correctly at leading order\footnote{In cases where relativistic sources on the right hand side of eq.~(\ref{eq:ij})
dominate over the second-order terms of the weak field expansion, the metric perturbations are still  determined correctly at leading order -- in this case the second-order contributions should be considered subleading. An explicit example will be given in Section \ref{sec:neutrino}.}, even in situations where they are strongly suppressed\footnote{Note, for instance, that our expansion agrees with a calculation in second-order perturbation theory of $\Lambda$CDM in the regime where the latter is valid. If we ignore primordial vector and tensor contributions,
the only first-order perturbations are scalars. At second order, $\chi$, $B_i$ and $h_{ij}$ are sourced by terms quadratic in the first-order
perturbations. We use the space-space components of Einstein's equations to determine $\chi$, $B_i$ and $h_{ij}$. To
acquire the correct number of spacetime indices, any term built from the linear scalar perturbations needs to have two spatial derivatives
in order to appear in this equation.
Our expansion therefore keeps all the relevant terms for a second-order calculation. Our stress-energy tensor, however, is computed
nonperturbatively. It coincides with the second-order one whenever second-order perturbation theory is valid,
but it remains valid also beyond this regime.}
(such as $\Lambda$CDM standard cosmology). In fact, this can be understood as the guiding principle of our weak field expansion: we construct a scheme which provides a meaningful
calculation of all the relativistic terms that perturb the geometry and are missed by a Newtonian treatment.

Applying the expansion to the time-time component of Einstein's equation, eq.~(\ref{eq:E00}), we obtain
\begin{equation}
 \label{eq:00}
 \left(1+4\Phi\right)\Delta\Phi - 3\cH\Phi' + 3\cH^2 (\chi- \Phi) + \frac{3}{2} \delta^{ij} \Phi_{,i} \Phi_{,j} = -4 \pi G a^2 \left(T^0_0 - \bar{T}^0_0\right) \, .
\end{equation}
If one wants to draw the analogy to a Newtonian scheme, this is the equation which replaces the Poisson equation $\Delta \psi_N = 4 \pi G a^2 \delta\rho$.
We use this equation to determine the evolution of $\Phi$. The remaining perturbation variables are determined from the traceless part of the space-space
set of Einstein's equations which have no Newtonian analog. In terms of the weak field expansion they read
\begin{multline}
\label{eq:ij}
 \left(\delta^i_k \delta^j_l - \frac{1}{3}\delta^{ij}\delta_{kl}\right)\biggl[\frac{1}{2}h''_{ij} + \cH h'_{ij} - \frac{1}{2}\Delta h_{ij} + B'_{(i,j)} + 2 \cH B_{(i,j)} + \chi_{,ij} - 2 \chi \Phi_{,ij} + 2 \Phi_{,i} \Phi_{,j} + 4 \Phi \Phi_{,ij}\biggr] \\
 = 8 \pi G a^2 \left(\delta_{ik} T^i_l - \frac{1}{3} \delta_{kl} T^i_i\right) \doteq 8 \pi G a^2 \Pi_{kl}\, ,
\end{multline}
where we introduced the anisotropic stress tensor  $\Pi_{kl}$. Due to the subtraction of the trace, these are five independent
equations, exactly the number needed in order to determine $\chi$ (scalar), $B_i$ (two polarizations) and $h_{ij}$ (two polarizations).

The remaining four Einstein's equations, namely the time-space equations and the spatial trace are redundant, they follow from the above equations and the covariant conservation
of stress-energy. The time-space equations are
\begin{equation}
\label{eq:0i}
 -\frac{1}{4}\Delta B_i - \Phi'_{,i} - \cH (\Phi_{,i}-\chi_{,i}) = 4 \pi G a^2 T^0_i \, .
\end{equation}
We  use this equation to verify that our code obtains consistent solutions.

The stress-energy tensor usually also depends on the metric variables, see eq.~(\ref{eq:genericTmunu}).
Therefore it is appropriate to expand the expression for the stress-energy tensor in terms of the metric perturbations. Truncating at linear order is consistent
with our weak field expansion if the dependence does not involve any derivatives of the metric. This should, however, not be confused with a perturbative treatment
of the entire stress-energy tensor. Instead, the fully nonlinear stress-energy is computed on a slightly perturbed geometry, and we therefore can ``dress'' the result by
geometric corrections in a perturbative way. An example of this procedure will be worked out in the next section.

To conclude the presentation of Einstein’s equations, table \ref{tab:weakfield} indicates the order of the different perturbation variables in our framework.
We strictly include all perturbations up to order $\epsilon$. The algorithms which solve eqs.~(\ref{eq:00}) and (\ref{eq:ij}) in our code are discussed
in Appendix~\ref{app:FFTsolvers}.

\begin{table}
\begin{center}
\begin{tabular}{|c|c|}
\hline
variable & order \\
\hline
$\Phi,~\Psi,~\Phi',~\Psi',~\Phi'',~\Psi''$ &  $\epsilon$ \\
$\Phi_{,i},~\Psi_{,i},~\Phi_{,i}',~\Psi_{,i}'$ &  $\epsilon^{1/2}$\\
$\Phi_{,ij},~\Psi_{,ij}$ &  $1$\\
$\chi,~\chi',~\chi'',~\chi_{,i},~\chi_{,i}',~\chi_{,ij}$ & $\epsilon$\\
$B_i,~B_i',~B_i'',~B_{i,j}~B_{i,j}'~B_{i,jk}$&   $\epsilon$\\
$h_{ij},~h_{ij}',~h_{ij}'',~h_{ij,k},~h_{ij,k}',~h_{ij,kl}$ &  $\epsilon$ \\
$\delta T^0_0 / \bar{T}^0_0$ & $1$ \\
$T^0_i/\bar{T}^0_0$ & $\epsilon^{1/2}$\\
$\Pi_{ij}/\bar{T}^0_0$ & $\epsilon$\\
$v^i,~q_i $ &$1$ \\
\hline
\end{tabular}
\end{center}
\caption{\label{tab:weakfield} The (most conservative) perturbative order assumed for various quantities in our framework.}
\end{table}

For a Universe dominated by relativistic sources (e.g.\ radiation or hot dark matter) we would have to consider $T^0_i / \bar{T}^0_0$ and $\Pi_{ij} / \bar{T}^0_0$ to be of the same order as $\delta T^0_0 / \bar{T}^0_0$. Indeed,
during radiation domination, all these quantities are of order $\epsilon$. However, in the late Universe relativistic sources are subdominant, and the gravitational instability of nonrelativistic matter leads to the above
hierarchy $\delta T^0_0 / \bar{T}^0_0 \sim 1 \gg T^0_i / \bar{T}^0_0 \sim \epsilon^{1/2} \gg \Pi_{ij} / \bar{T}^0_0 \sim \epsilon$. This follows from the fact that for cold dark matter the velocities
are $\sim \Psi_{,i} \sim \epsilon^{1/2}$. For relativistic particles the velocities are order unity, but these particles do not cluster and therefore maintain a $\delta T^0_0 / \bar{T}^0_0$ of order
$\epsilon$. The individual particle velocities or momenta, $v^i,~q_i$ introduced in the next section, are treated fully relativistically in order to cover both, relativistic and
nonrelativistic matter.

\section{The particle ensemble}
\label{sec:particles}

Particles are a possible source for stress-energy perturbations relevant for many applications. This includes standard Cold Dark Matter (CDM) and baryons, which are non-relativistic
during the nonlinear stage of structure formation, but also relativistic species such as light neutrinos or Warm Dark Matter (WDM). Newtonian N-body codes are suitable for non\-rela\-tivistic
particles, since their system of equations relies on the fact that velocities are much smaller than the speed of light. Nevertheless, simulations with neutrinos or WDM have been
carried out with such codes \cite{Bode:2000gq,VillaescusaNavarro:2012ag,Inman:2015pfa}. Such simulations are often initialized at a time when most of the particles have become non-relativistic,
setting up an initial distribution which takes into account some aspects of relativistic early evolution. Adhering to our relativistic approach, we do not make any assumptions about
the distribution in momentum space, allowing for arbitrarily high momenta.

\subsection{Relativistic momentum and geodesic equations}
\label{sec:momentum}

To set up our relativistic description we  start with the classical action of a massive point-particle,
\begin{equation}
\label{eq:1p-action}
 \mathcal{S}_\mathrm{p} = -m \int\! \vert \dd s \vert = -m \int\! \sqrt{-g_{\mu\nu} \frac{\dd x^\mu}{\dd\tau} \frac{\dd x^\nu}{\dd\tau}} \dd\tau \doteq \int\! \mathcal{L}_\mathrm{p} \dd\tau \, .
\end{equation}
At linear order in the metric perturbations, the Lagrangian $\mathcal{L}_\mathrm{p}$ can be written as
\begin{equation}
 \mathcal{L}_\mathrm{p} = -m a \sqrt{1-v^2} \left(1 + \frac{\Psi + v^2 \Phi + B_i v^i - \frac{1}{2} h_{ij} v^i v^j}{1-v^2}\right) \, ,
\end{equation}
where $v^i \doteq \dd x^i / \dd\tau$ is the coordinate three-velocity and $v^2 \doteq \delta_{ij} v^i v^j$. In order to simplify notation we define $v_i \doteq \delta_{ij} v^j$ as the
symbol $v_i$ with lower index is not otherwise used.

The canonical conjugate momentum to $x^i$, defined as $q_i \doteq \partial \mathcal{L}_\mathrm{p} / \partial v^i$, then reads
\begin{equation}
 q_i = \frac{m a}{\sqrt{1-v^2}} \left[v_i \left(1-2\Phi-\frac{\Psi + v^2 \Phi + B_j v^j - \frac{1}{2} h_{jk} v^j v^k}{1-v^2}\right) - B_i + h_{ij} v^j\right] \, ,
\end{equation}
and again, we define $q^i \doteq \delta^{ij} q_j.$ We invert this expression at linear order in the metric perturbations, 
\begin{equation}
\label{eq:v_of_q}
 v_i = \frac{q_i}{\sqrt{q^2 + m^2 a^2}} \left[1 + \Psi + \left(2 - \frac{q^2}{q^2 + m^2 a^2}\right)\! \Phi + \frac{1}{2} \frac{q^j q^k  h_{jk}}{q^2 + m^2 a^2}\right] + B_i - \frac{h_{ij} q^j}{\sqrt{q^2 + m^2 a^2}}\, ,
\end{equation}
where $q^2 \doteq \delta^{ij} q_i q_j$. The Euler-Lagrange equation, $\dd q_i / \dd\tau = \partial \mathcal{L}_\mathrm{p} / \partial x^i$, can then be written as
\begin{equation}
\label{eq:geodesic}
 \frac{\dd q_i}{\dd\tau} = -\sqrt{q^2 + m^2 a^2} \left(\Psi_{,i} + \frac{q^2}{q^2 + m^2 a^2} \Phi_{,i} + \frac{q^{j} B_{j,i}}{\sqrt{q^2 + m^2 a^2}} - \frac{1}{2} \frac{q^j q^k  h_{jk,i}}{q^2 + m^2 a^2}\right) \, ,
\end{equation}
where we used eq.~(\ref{eq:v_of_q}) to replace $v^i$. The last two equations are the geodesic equations for a massive particle (carrying arbitrary momentum) in
a linearly perturbed geometry. Since only first derivatives of the metric appear, they contain all the necessary terms to be consistent with our weak field expansion.
The Newtonian limit is obtained when $q^2 \ll m^2 a^2$ and only the first term of eq.~(\ref{eq:geodesic}) is retained. A similar derivation was presented in \cite{Ma:1993xs} for the case where only scalar perturbations are present.

Since eq.~(\ref{eq:geodesic}) is much simpler than a corresponding equation for $\dd v^i / \dd\tau$ like the one presented in \cite{Adamek:2014xba}, our code 
directly evolves $q_i$, from which $v^i$ can always be recovered using eq.~(\ref{eq:v_of_q}). In the next subsection, we will derive expressions for the stress-energy
tensor of an ensemble of point-particles in terms of their canonical momenta.

\subsection{Stress-energy tensor}
\label{sec:stress-energy}

The action of an ensemble of point-particles is given by the sum over the one-particle actions (\ref{eq:1p-action}),
\begin{equation}
 \mathcal{S}_\mathrm{m} = \int\! \sqrt{-g} \mathcal{L}_\mathrm{m} \dd^4 x = \sum_n \int\! \delta^{(3)}(\mathbf{x}-\mathbf{x}_{(n)}) \mathcal{L}_\mathrm{p}[x^i_{(n)}, v^i_{(n)}, \tau] \dd^3\mathbf{x} \dd\tau \, ,
\end{equation}
where $\mathbf{x}_{(n)}$ or $x^i_{(n)}$ both denote the position of the $n$th particle, and $v^i_{(n)}$ is the respective coordinate three-velocity.
We can use eq.~(\ref{eq:genericTmunu}) to obtain the stress-energy tensor from this action. Expanding again in terms of the metric perturbation variables and
using eq.~(\ref{eq:v_of_q}) to replace $v^i_{(n)}$ in favour of $q_i^{(n)}$, the canonical momentum for the $n$th particle, we find following expressions
for the components of the stress-energy tensor, $T^\mu_\nu$:
\begin{align}
\label{eq:T00}
 T^0_0 = & -\frac{1}{a^4} \sum_n \delta^{(3)}(\mathbf{x}-\mathbf{x}_{(n)}) \Biggl[\sqrt{q^2_{(n)} + m^2_{(n)} a^2} \left(1 + 3 \Phi + \frac{q^2_{(n)}}{q^2_{(n)} + m^2_{(n)} a^2} \Phi\right)\Biggr. \nonumber\\
 &\hspace{7.5cm}\Biggl.+ q^i_{(n)} B_i - \frac{q^i_{(n)} h_{ij} q^j_{(n)}}{2 \sqrt{q^2_{(n)} + m^2_{(n)} a^2}} \Biggr]\, ,\\
 \label{eq:Tij}
 T^i_j = & \frac{\delta^{ik}}{a^4} \sum_n \delta^{(3)}(\mathbf{x}-\mathbf{x}_{(n)}) \Biggl[\frac{q_j^{(n)} q_k^{(n)}}{\sqrt{q^2_{(n)} + m^2_{(n)} a^2}} \left(1 + 4 \Phi + \frac{m^2_{(n)} a^2}{q^2_{(n)} + m^2_{(n)} a^2} \Phi + \frac{h_{lm} q^l_{(n)} q^m_{(n)}}{2 \left(q^2_{(n)} + m^2_{(n)} a^2\right)}\right)\Biggr. \nonumber\\
 &\hspace{7.5cm}\Biggl.+ q_j^{(n)} B_k - \frac{q_j^{(n)} h_{kl} q^l_{(n)}}{\sqrt{q^2_{(n)} + m^2_{(n)} a^2}} \Biggr]\, ,\\
 \label{eq:Ti0}
 T^i_0 = & -\frac{\delta^{ij}}{a^4} \sum_n \delta^{(3)}(\mathbf{x}-\mathbf{x}_{(n)}) \Biggl[ q_j^{(n)} \left(1 + 5 \Phi + \Psi + \frac{q^l_{(n)} B_l}{\sqrt{q^2_{(n)} + m^2_{(n)} a^2}}\right)\Biggr. \nonumber\\
 &\hspace{7.5cm}\Biggr.+ \sqrt{q^2_{(n)} + m^2_{(n)} a^2} B_j - q^k_{(n)} h_{jk}\Biggr]\, ,\\
 \label{eq:T0i}
 T^0_i = & \frac{1}{a^4} \sum_n \delta^{(3)}(\mathbf{x}-\mathbf{x}_{(n)}) q_i^{(n)} \left(1 + 3 \Phi - \Psi\right) \, .
\end{align}
From eqs.~(\ref{eq:0i}) and (\ref{eq:ij}) we note that $T^0_i$ and the traceless part of $T^i_j$ are first order quantities.
Therefore, according to our counting scheme, terms like
\begin{equation}
 \frac{1}{a^4} \sum_n \delta^{(3)}(\mathbf{x}-\mathbf{x}_{(n)}) B_i q_{j}^{(n)} \simeq B_i T^0_j \, ,
\end{equation}
\begin{equation}
\frac{1}{a^4} \sum_n \delta^{(3)}(\mathbf{x}-\mathbf{x}_{(n)}) h_{ij} q_{k}^{(n)} \simeq h_{ij} T^0_k \, ,
\end{equation}
and
\begin{equation}
 \frac{1}{a^4} \sum_n \delta^{(3)}(\mathbf{x}-\mathbf{x}_{(n)}) \frac{h_{ij} q^i_{(n)} q^j_{(n)}}{\sqrt{q^2_{(n)} + m^2_{(n)} a^2}} \simeq h_{ij} \delta^{ik} \delta^{jl} \Pi_{kl}
\end{equation}
are higher order in our weak field expansion and can safely be neglected. The second line of eq.~(\ref{eq:T00}) can then be dropped completely.
The second line of eq.~(\ref{eq:Tij}), on the other hand, simplifies to
\begin{equation}
 \frac{1}{a^4} \sum_n \delta^{(3)}(\mathbf{x}-\mathbf{x}_{(n)}) \Biggl[q_j^{(n)} B_k - \frac{q_j^{(n)} h_{kl} q^l_{(n)}}{\sqrt{q^2_{(n)} + m^2_{(n)} a^2}} \Biggr] \simeq -h_{jk} \bar{P} \, ,
\end{equation}
where $\bar{P} \doteq \bar{T}^i_i / 3$ is the isotropic pressure in the background model of the N-body ensemble, and the second
line of eq.~(\ref{eq:Ti0}) finally becomes
\begin{equation}
 -\frac{1}{a^4} \sum_n \delta^{(3)}(\mathbf{x}-\mathbf{x}_{(n)}) \biggl[\sqrt{q^2_{(n)} + m^2_{(n)} a^2} B_j - q^k_{(n)} h_{jk}\biggr] \simeq B_j T^0_0\, .
\end{equation}

\section{Code structure}
\label{sec:code}

Here we give an overview of the various elements of our new N-body code, called \textit{gevolution}, and how they
work together in order to solve the coupled dynamics of metric and matter degrees of freedom. Some details mainly relevant for users
who want to modify our code for their purposes (for instance by adding new degrees of freedom to the matter Lagrangian) are
relegated to several technical appendices.

The basic design concept of the code is a particle-mesh scheme. This means that a spacelike hypersurface is tessellated using a
regular Cartesian lattice. Any continuous fields, for instance the metric perturbations or the stress-energy tensor, are discretized
by sampling their values on this lattice, i.e.\ at the coordinates of lattice points. Like many Newtonian codes we use periodic boundary
conditions, meaning that the global topology of the hypersurface is chosen\footnote{This choice is not apparent at the level of
the dynamics which is described by local equations, but it places a constraint on the state of the system. We are not aware of any procedure
to avoid such a constraint within a numerical scheme. It is, however, important to take this into account, for instance when discussing
correlation functions.} to be a three-torus or, equivalently, an infinite Euclidean space obtained by periodically repeating the exact
perturbation pattern of a single cubic template. The partial differential equations (\ref{eq:00}) and (\ref{eq:ij}) which determine
the metric perturbations are solved on the lattice by replacing the differential operators by finite-difference versions thereof.
As explained in Appendix \ref{app:FFTsolvers}, the current implementation uses Fourier analysis to solve all these finite-difference equations.

The second element of the particle-mesh scheme is the ensemble of N-body particles, which is the method of choice for discretizing the particle phase space. Because of its dimensionality it is unfeasible to employ a lattice discretization on phase space. Instead, one discretizes the distribution function by drawing a representative sample. Each N-body particle can be considered as a discrete element of phase space, following its phase flow as it evolves with time.

The positions and momenta of N-body particles can take arbitrary values, meaning that they exist independently of the lattice. However, the evolution of fields and particles mutually depends on one another. The relations are established by means of projection and interpolation methods as explained in Appendix \ref{app:PM}. For instance, the stress-energy tensor on the lattice is obtained by a particle-to-mesh projection. Vice versa, in order to solve the geodesic equations, the metric fields are interpolated to the particle positions.

The code \textit{gevolution} is built on top of the \lf\ library \cite{David:2015eya}. This library has been originally developed to simplify the implementation of classical lattice based field simulations on massively parallel computers.
\lf , written in C++, manages the parallelization using the MPI protocol. The work is distributed based on a rod-decomposition of
the lattice. The library also provides a parallel implementation of the three-dimensional Fast Fourier Transform (FFT) which is scalable up to very large numbers of MPI processes. For \textit{gevolution}, \lf\ has been extended to handle particle lists (with arbitrary properties) and their mapping onto the lattices.

Schematically, the main loop of \textit{gevolution} consists (at least) of the following steps:
\begin{enumerate}
 \item Compute the new momenta for the particles using eq.~(\ref{eq:geodesic}).
 \item Evolve the background by half a time step.
 \item Update the particle positions using eq.~(\ref{eq:v_of_q}).
 \item Evolve the background by half a time step; background and particles are now at the new values while metric perturbations
 still require update.
 \item Construct $T^0_0$ and compute the terms quadratic in $\Phi$  and linear in $\chi$ of eq.~(\ref{eq:00}) using the present values.
 The new value of $\Phi$ is then determined by solving a linear diffusion equation (with nonlinear source) using Fourier analysis.
 \item Construct $T^i_j$ and compute the terms quadratic in $\Phi$ of eq.~(\ref{eq:ij}) using the new value of $\Phi$.
 After moving again to Fourier space the equation can be separated into its various spin components.
 The spin-0 projection determines the new value of $\chi$ via an elliptic constraint. The spin-1 projection gives a parabolic evolution
 equation for $B_i$, whereas the spin-2 projection is a hyperbolic wave-equation for $h_{ij}$. All metric perturbations are now at the
 new values; time step complete.
\end{enumerate}
This list assumes an ensemble of nonrelativistic N-body particles as the only source of matter. If other matter degrees of freedom are
present, appropriate update steps need to be added in order to integrate their dynamics. An example will be discussed in Section
\ref{sec:neutrino}.

The time stepping is chosen as small as necessary for numerical convergence. It therefore depends on the convergence properties of the
numerical solvers used for the matter evolution. For nonrelativistic matter a typical criterion would be to require that particles do
not travel farther than one lattice unit in each update.

Apart from the main loop which solves the dynamics, \textit{gevolution} has some routines which are responsible for initialization and output.
Initial conditions can be generated ``on the fly'' which is usually faster than reading them from disk. Some details are
given in Appendix \ref{app:ICs}. An interface with \textit{FalconIC} \cite{Valkenburg:2015dsa} has also been implemented, allowing to call
various Boltzmann codes directly at runtime in order to generate initial data. Output consists either of so-called ``snapshots'', i.e.\ the three dimensional representation of fields
and particles at a specific coordinate time, or of power spectra. As usual, the latter represent a very convenient form of data reduction.

\section{Simulation results}
\label{sec:simulations}

In this section we show some first results obtained with our new code.

\subsection{Isolated point-mass}
\label{sec:pointmass}

Due to the periodic boundary conditions, a single mass placed inside our simulation volume  represents a regular lattice of point masses. However, in the close vicinity of the
point-mass it is justified to neglect the influence of the other far-away masses, just as one can neglect the influence of other stars when studying the orbits of planets
around the sun. Hence we expect that for a region much smaller than the simulation volume, the metric around a point-mass should agree with the Schwarzschild solution.
This setup can therefore be used to validate the numerical solvers for the relativistic potentials $\Psi$ and $\Phi$ in vacuum, i.e.\ independently of matter dynamics.

\begin{figure}[t]
 \centerline{\includegraphics[width=0.62\textwidth]{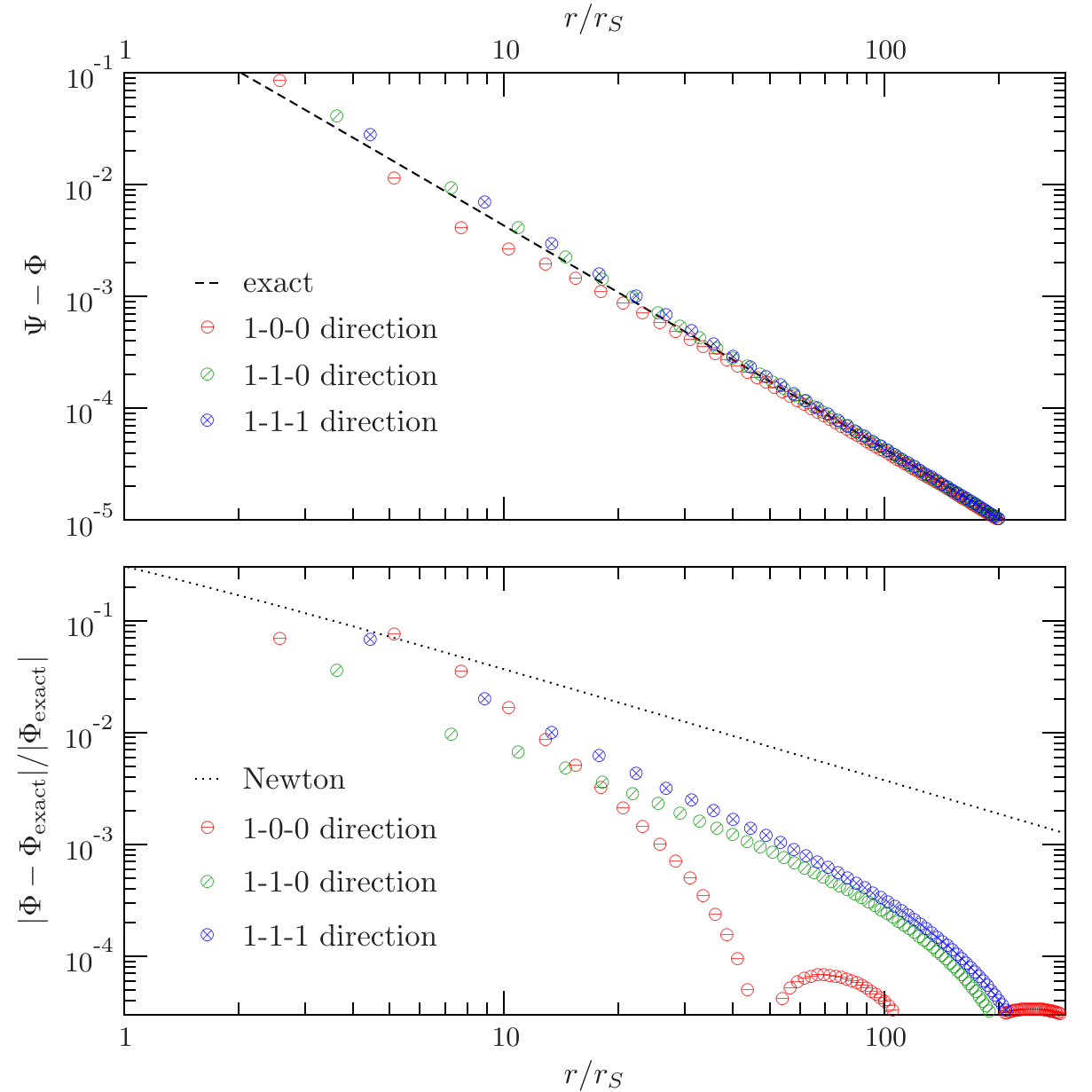}}
\caption{\label{fig:Schwarzschild} We compare the numerical values of the relativistic potentials $\Psi$ and $\Phi$ around a single point-mass with the analytic expressions given by the Schwarzschild solution.
Since the lattice breaks isotropy, we sample the potentials along three different lattice directions. The top panel shows the difference between the two potentials, $\Psi-\Phi$, which
is zero in the Newtonian approximation, i.e.\ it goes as $(r_S/r)^2$ in the weak field regime. The bottom panel shows the relative error of one of the two potentials, $\Phi$, which
is computed at one order better than Newtonian (shown as dotted line). In other words, both $\Phi$ and $\Psi$ are solved correctly up to order $(r_S/r)^3$-terms, as opposed to the
Newtonian limit where terms of order $(r_S/r)^2$ are neglected as well. We notice that our numerical errors show some dependence on direction, but are always of the same order in the
weak field expansion. The simulation was set up on a lattice with $6144^3$ points, which means that  finite volume effects kick in at distances $r \gtrsim 1000$ times the lattice unit.
We cut off our plots well before this becomes an issue.}
\end{figure}

In Figure~\ref{fig:Schwarzschild} we show the relativistic potentials around a single particle. The coarse-graining introduced by the particle-mesh
scheme effectively distributes the mass over one lattice cell, which was in this case chosen to be somewhat larger than the Schwarzschild radius in order to
enforce weak field conditions for all the scales resolved by the simulation. In addition, the discrete symmetries of the lattice break isotropy. We can study
this effect by comparing the potentials along different directions. The values of the potentials are plotted as function of the distance from the point-mass
for three different lattice directions, 1-0-0, 1-1-0, and 1-1-1. 

In order to compare the numerical results with the Schwarzschild solution, we write the Schwarzschild metric using so-called ``isotropic coordinates'' \cite{Eddington},
\begin{equation}
\label{eq:Schwarzschild}
 \dd s^2 = -\frac{\left(1-\frac{r_S}{4r}\right)^2}{\left(1+\frac{r_S}{4r}\right)^2} \dd t^2 + \left(1+\frac{r_S}{4r}\right)^4\left[\dd r^2 + r^2 \dd \Omega^2\right] \, ,
\end{equation}
which, by comparing with eq.~(\ref{eq:metric}), gives following analytic expressions for the relativistic potentials as a function of the coordinate distance $r$:
\begin{align}
 \Psi_\mathrm{exact}(r) = &\, \frac{1}{2}\frac{\left(1-\frac{r_S}{4r}\right)^2}{\left(1+\frac{r_S}{4r}\right)^2}-\frac{1}{2} \, ,\\
 \Phi_\mathrm{exact}(r) = &\, \frac{1}{2}-\frac{1}{2}\left(1+\frac{r_S}{4r}\right)^4 \, .
\end{align}
Here, $r_S \doteq 2 G M$ denotes the Schwarzschild radius, and we note $B_i = h_{ij} = 0$ in the Schwarzschild solution. We should also point out that, for the purpose of
this comparison, we use Minkowski space as the background model since the entire simulation volume is in vacuum except for a single cell. In other words, we set $\bar{T}^\mu_\nu = 0$
which implies $\cH = 0$ by eq.~(\ref{eq:Friedmann}), and therefore we can set $a=1$ and identify $t=\tau$.

It is instructive to expand the exact expressions for $r \gg r_S$, which gives
\begin{align}
 \Psi(r) = &\, -\frac{r_S}{2r} + \frac{r_S^2}{4r^2} - \frac{3 r_S^3}{32 r^3} + \ldots \, ,\\
 \Phi(r) = &\, -\frac{r_S}{2r} - \frac{3r_S^2}{16r^2} - \frac{r_S^3}{32 r^3} - \frac{r_S^4}{512 r^4} \, .
\end{align}
The Newtonian limit is given by the first term in this expansion, $\Psi = \Phi = -r_S/2r$. The next weak field correction to this is suppressed by another power of $r_S/r$, and so on.
As is evident from Figure~\ref{fig:Schwarzschild}, our numerical scheme correctly accounts for the next-to-leading order terms, meaning that the errors are suppressed by an additional
power of $r_S/r$ as compared to the Newtonian approximation. We note in passing that these are the terms relevant for the perihelion advance
of Mercury. A similar study was presented in \cite{Adamek:2015hqa}, where our framework is applied to a spherically symmetric setup.

We want to emphasize that the accuracy discussed here is a fundamental limitation of the weak field expansion which is independent
of the discretization scheme. We expect our weak field approximation to break down at next-to-next-to-leading order (in the above expansion)
even at infinite resolution. In practice, however, one has to deal with additional discretization errors. For the numerical test presented
here we chose an extremely high resolution such that these effects are subdominant. This generally is not possible in many practical
applications.

Figure \ref{fig:Schwarzschild} shows that our code models the space-time geometry around a Schwarzschild black hole correctly and with sub-percent precision for $r \gtrsim 10\, r_S$. In other words, our weak field limit is accurate as long as our resolution is larger than about 0.1 parsec, as the Schwarzschild radius of a $10^{11} M_\odot$ black hole (about 5 times larger than the largest currently known black hole) is only about 0.01 pc. The same is not true for Newtonian codes as e.g.\ $\Phi-\Psi$ is identically zero in this approximation. Of course on such small scales many other effects that are currently neglected, like baryons, play an important role.

\subsection{Code comparison with Gadget-2 and RAMSES}
\label{sec:gadget}

In order to test further some aspects of our implementation, in particular the particle-mesh scheme, we added the option to run
simulations using Newton's theory of gravity in our code. With this option, instead of the full stress-energy tensor, the code
computes a Newtonian density contrast $\delta_N$ by particle-mesh projection of the rest-mass distribution. Then, a Newtonian
potential $\psi_N$ is computed by solving
\begin{equation}
\label{eq:Poisson}
 \Delta \psi_N = 4 \pi G a^2 \bar{\rho} \delta_N \, ,
\end{equation}
using Fourier analysis. Here $\bar{\rho}$ is the average rest-mass density which, in a Newtonian setting, coincides with the background model
for $-\bar{T}^0_0$ of matter. The algorithm used for eq.~(\ref{eq:Poisson}) is fundamentally very similar to the one which is used for
solving eq.~(\ref{eq:00}). Finally, for the geodesic equation we take the Newtonian limit where
\begin{eqnarray}
\label{eq:Newtonvel}
 v_i &=& \frac{q_i}{m a} \, ,\\
\label{eq:Newtonforce}
 \frac{\dd q_i}{\dd\tau} &=& -m a \left(\psi_N\right)_{,i} \, .
\end{eqnarray}
With these modifications the code therefore solves the Newtonian system of equations using essentially the same numerical techniques
which we want to employ for the relativistic simulations. Of course the relativistic framework is much richer and needs several additional
algorithms, but at least some parts of the implementation can now be compared to existing codes.

\begin{figure}[t]
 \centerline{\includegraphics[width=\textwidth]{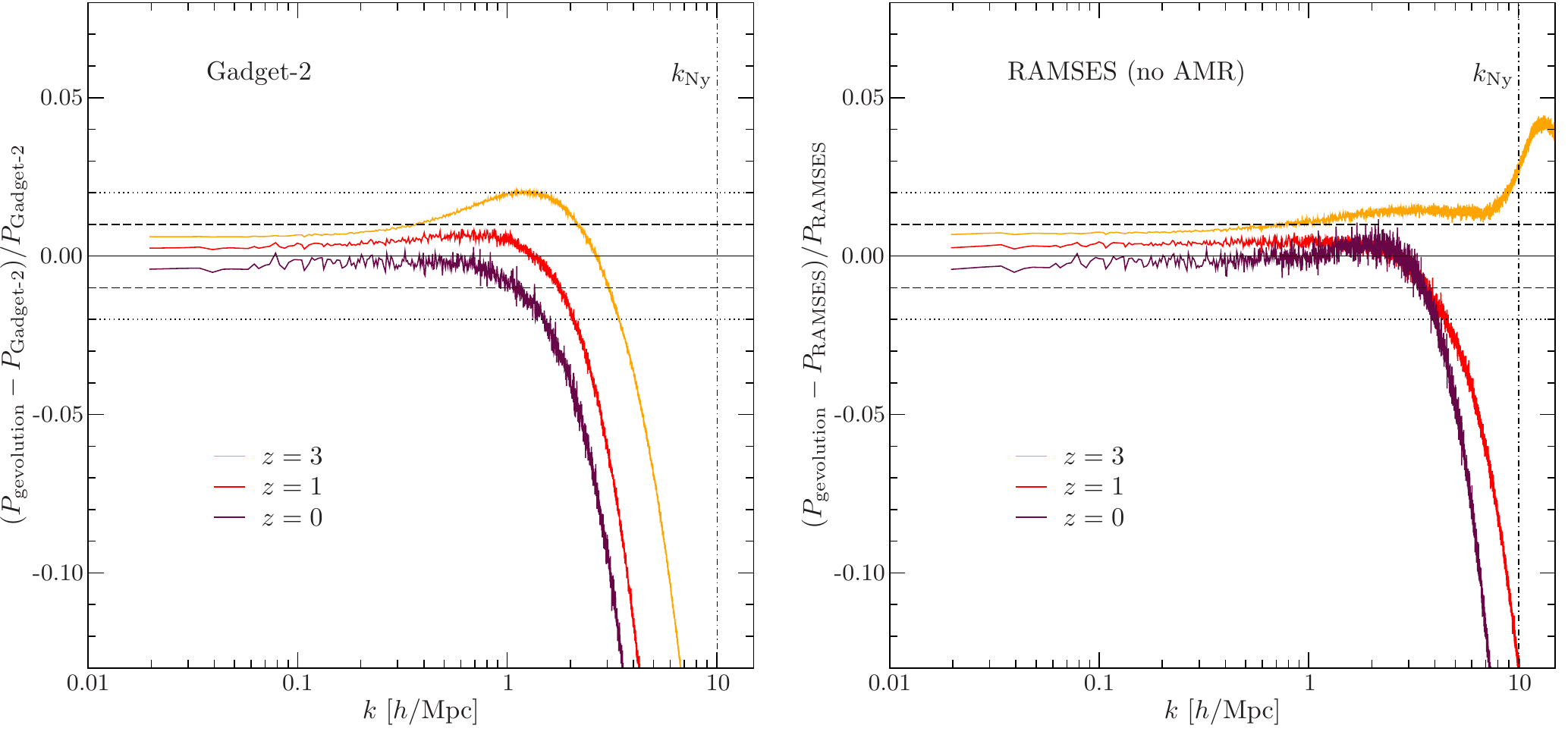}}
\caption{\label{fig:gadget} We show the relative difference of the matter power spectra of snapshots from three Newtonian
simulations which
were initialized on identical particle configurations. One simulation was done with \textit{gevolution} and ``gravity theory'' set
to ``Newton'', one was done with \textit{Gadget-2}, and the third one was done with \textit{RAMSES}. The left panel shows the comparison
of \textit{gevolution} versus \textit{Gadget-2}, the right panel shows the one of \textit{gevolution} versus \textit{RAMSES}. Since our code works at fixed resolution, characterized by a
Nyqvist frequency $k_\mathrm{Ny}$ (vertical dot-dashed line), it is expected that it can not resolve the formation of small scale
structures with $k \gtrsim k_\mathrm{Ny}$. The same is true for the RAMSES simulation since we did not allow the code to perform
any adaptive mesh refinement (AMR). Hence the agreement is better in this case but differences in the discretization scheme still lead
to sizable effects as one approaches $k_\mathrm{Ny}$. However, on scales $k \ll k_\mathrm{Ny}$ all simulations agree to better than one
percent, even in the nonlinear regime.}
\end{figure}

\begin{figure}[t]
 \centerline{\includegraphics[width=0.62\textwidth]{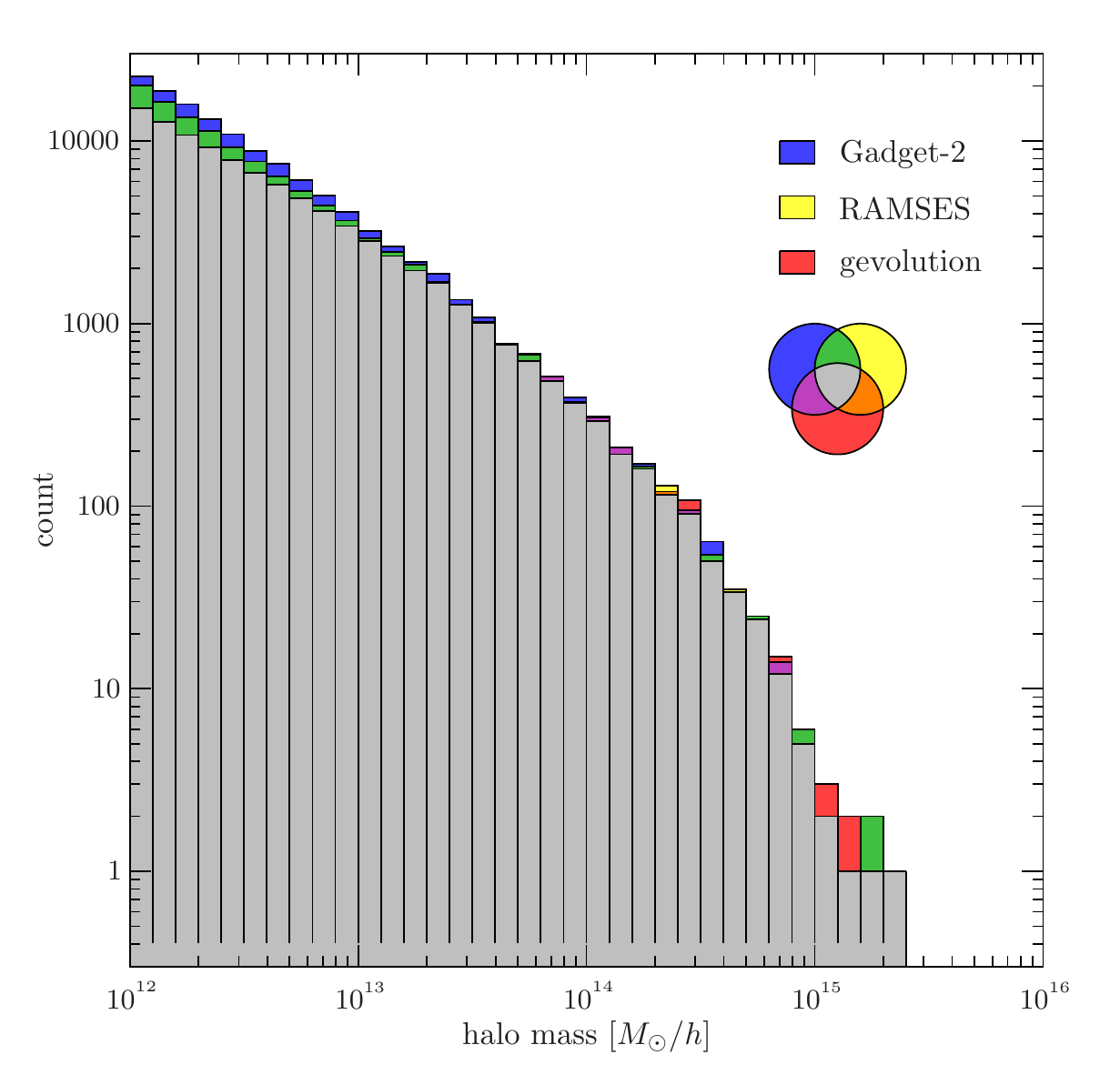}}
\caption{\label{fig:hmf} For the same simulations as in Figure \ref{fig:gadget}, we show the halo mass distributions found at
redshift $z = 0$. The lack of small scale structure in \textit{gevolution} relative to \textit{Gadget-2} due to the fixed spatial resolution
of the former leads to a deficit of small mass halos. The effect can also be seen for the simulation done with \textit{RAMSES} which
used the same fixed resolution. The deficit is a bit smaller in this case, possibly due to the fact that \textit{RAMSES} uses a higher-order
force interpolation. The regime of large masses shows excellent agreement between all three simulations.}
\end{figure}

We choose to compare our implementation with \textit{Gadget-2} \cite{Springel:2000yr,Springel:2005mi} and with \textit{RAMSES} (version 3.0) \cite{Teyssier:2001cp}. The numerical approach used in \textit{Gadget-2} is very different from ours, in particular with respect to the force
computation at short distances. The agreement of the results  hence is a good indication for the robustness of the N-body approach
and allows to assess the detriment of working at fixed resolution. \textit{RAMSES} on the other hand, like \textit{gevolution}, uses a particle-mesh scheme but it can increase force resolution
by performing adaptive mesh refinement. However, for our comparison we disable this feature, forcing the code to work at fixed resolution. This means we are looking at two cases: one where the numerical schemes are very different and one where they are quite similar (but still different).

We generate an
initial particle configuration corresponding to a $\Lambda$CDM cosmology\footnote{As explained in Appendix \ref{app:ICs} the linear
particle displacement for a Newtonian simulation differs from the one of Poisson gauge used for our relativistic simulations. This
issue is taken into account when we generate initial data.} which is then evolved by all three codes.
Using identical initial data the comparison is not affected by cosmic variance. Figure \ref{fig:gadget} shows the relative difference of the
power spectra after the codes have evolved the particle configuration from initial redshift $z_\mathrm{in} = 63$ to $z = 3$, $1$, $0$.

The left panel shows the comparison between \textit{Gadget-2} and \textit{gevolution}. As may be expected, our code displays a significant deficit in power as one approaches the Nyqvist frequency. This is due to the fact that we work at
fixed resolution. On scales well below the Nyqvist frequency, however, the results agree to within one 
percent, even in the nonlinear regime. It should be noted that an even better agreement would be hard to achieve since both codes employ a
first-order-in-time integration scheme but are based on very different numerical approaches. In other words, the 
difference is roughly what one expects for two different integration methods (see also \cite{Schneider:2015yka}).

The right panel shows the comparison between \textit{RAMSES} and \textit{gevolution}. As both simulations were run at the same fixed
resolution the agreement is significantly better. However, the discretization schemes are slightly different, in particular concerning the force interpolation (\textit{RAMSES} uses a five-point gradient) and the time integration (\textit{RAMSES} works in ``super-conformal time'').
Because of the former there is no reason why the codes should agree as one approaches the Nyqvist frequency. The latter, on the other hand,
may be responsible for the small constant offset at large scales which however remains within one percent.

Figure \ref{fig:hmf} shows the comparison of the halo mass distributions of DM halos at redshift $z = 0$, extracted for the
simulations using the \textit{Rockstar} halo finder \cite{Behroozi:2011ju}. As expected from the preceding discussion, the
agreement on the count of large-mass halos is very good, while \textit{gevolution} shows a deficit of small-mass objects. Nevertheless,
at the resolution used for this comparison, the halo abundance found with \textit{gevolution} appears to be reliable over two orders of
magnitude in mass.

The comparison shown in Figures \ref{fig:gadget} and \ref{fig:hmf} was based on simulations with $1024^3$ particles
and a comoving box of $320$ Mpc/$h$, which results in a mass resolution of $\sim 2.64\times10^9~M_\odot/h$. The lattice in \textit{gevolution} and \textit{RAMSES} had $1024^3$ points, whereas \textit{Gadget-2} used
a $512^3$ lattice for computing the long-range forces and its tree algorithm for the short-range contributions. The latter allows
\textit{Gadget-2} to resolve much smaller scales\footnote{Note, however, that there is no initial power beyond the Nyqvist frequency
of the particle ensemble.} and therefore requires significantly more work. We found that the simulation done with \textit{Gadget-2} consumed
about seven times as many CPU-hours as the simulation done with our code, even though both codes integrated for roughly the same number of
time steps. Restoring the relativistic setting some of this advantage is lost due to the additional work required for solving a more
complicated geodesic equation. The additional metric components $\chi$ and $B_i$ are also 
computed and recorded in a Newtonian run of our code --- they are just 
not used in the dynamics. For this reason we find that \textit{RAMSES} (without adaptive mesh refinement) runs about twice as fast as \textit{gevolution}. We nonetheless demonstrate that our relativistic scheme is not fundamentally much more expensive than a
Newtonian one.

Having established good agreement with existing numerical codes when working in a Newtonian setting one can proceed to study all
the aspects of the relativistic framework by \textit{internal} comparison. In other words, by switching between the two
theories of gravity (Newton and GR) within our code one can study the differences based on a single implementation without
changing the systematics, the data format, or other collateral elements which complicate matters.

\subsection{Background evolution}
\label{sec:backreaction}

In order to provide further insight into the role of the background model, elucidating the reasoning of
Section \ref{sec:background}, we conduct following numerical study. Let us assume we want to simulate a cosmology with
massive neutrinos where the neutrino phase space is
 sampled with the N-body method. While the details of such simulations are the topic of the next section,
here we are only concerned with the background model. Of course the usual way to construct the background model
would be to integrate over the unperturbed phase space distribution functions of the neutrino species in order to determine
their background energy densities which enter Friedmann's equation. This is, in fact, how the background model used in the
next section is constructed. It is also the background model assumed by linear Boltzmann codes such as \textit{CAMB} \cite{Lewis:1999bs}
or \textit{CLASS} \cite{Blas:2011rf}.

Let us compare this background model with an even simpler approximation where we completely ignore the kinetic energy of the
neutrinos and only consider their rest mass contribution to the background. This is a good approximation as soon as the
neutrinos are non-relativistic, but it  leads to some noticeable differences at higher redshifts\footnote{In the radiation dominated
era this approximate background model would, in fact, be terribly wrong. However, here we want to use the approximate model only during
N-body evolution, i.e.\ long after the end of radiation domination.}. Since we argue that the choice of background model is to some extent arbitrary, we want to understand what happens if one uses this approximate background model in our N-body code.

As should be clear from the discussion in Section \ref{sec:background}, the subtraction of an imperfect background model in
eq.~(\ref{eq:E00})  leads to a homogeneous mode in this equation, i.e.\ to the appearance of a non-vanishing homogeneous mode in
the scalar metric perturbations $\Phi$ and $\Psi$. Let us denote this homogeneous mode as $\bar{\Phi}$, and we remind the reader that we use
our gauge freedom to set $\bar{\Psi} = \bar{\Phi}$, which corresponds to a certain choice of global time coordinate $\tau$. The spatially
homogeneous part of the metric is then characterized by following line element:
\begin{equation}
\label{eq:approximatemodel}
 \dd s^2 = a^2(\tau) \left[-\left(1+2\bar{\Phi}\right) \dd\tau^2 + \left(1-2\bar{\Phi}\right) \delta_{ij} \dd x^i \dd x^j\right]
\end{equation}
Furthermore, the approximate background model has a certain conformal Hubble function $\cH(z)$ determined by solving the Friedmann equation for that model, where $z \doteq (1/a) - 1$.

Notice, however, that we can define the following \textit{renormalized} scale factor $\tilde{a} \doteq a (1 - \bar{\Phi})$ and \textit{renormalized} time coordinate $\dd \tilde{\tau} \doteq \dd \tau (1 + 2 \bar{\Phi})$, such that the new line element looks like an unperturbed Friedmann model,
\begin{equation}
\label{eq:renormalizedmodel}
 \dd s^2 = \tilde{a}^2(\tilde{\tau}) \left[-\dd \tilde{\tau}^2 + \delta_{ij} \dd x^i \dd x^j\right] \, ,
\end{equation}
up to higher-order corrections which we neglect in our weak-field expansion. The \textit{renormalized} Hubble function is, to leading
order in the weak-field expansion, related to the original one as
\begin{equation}
\label{eq:Hren}
\tilde{\cH} = \cH \left(1 - 2 \bar{\Phi} - a \frac{\dd \bar{\Phi}}{\dd a}\right) \, , \qquad \tilde{z} + 1 = \left(z + 1\right) \left(1 + \bar{\Phi}\right) \, . 
\end{equation}

\begin{figure}[t]
 \centerline{\includegraphics[width=0.62\textwidth]{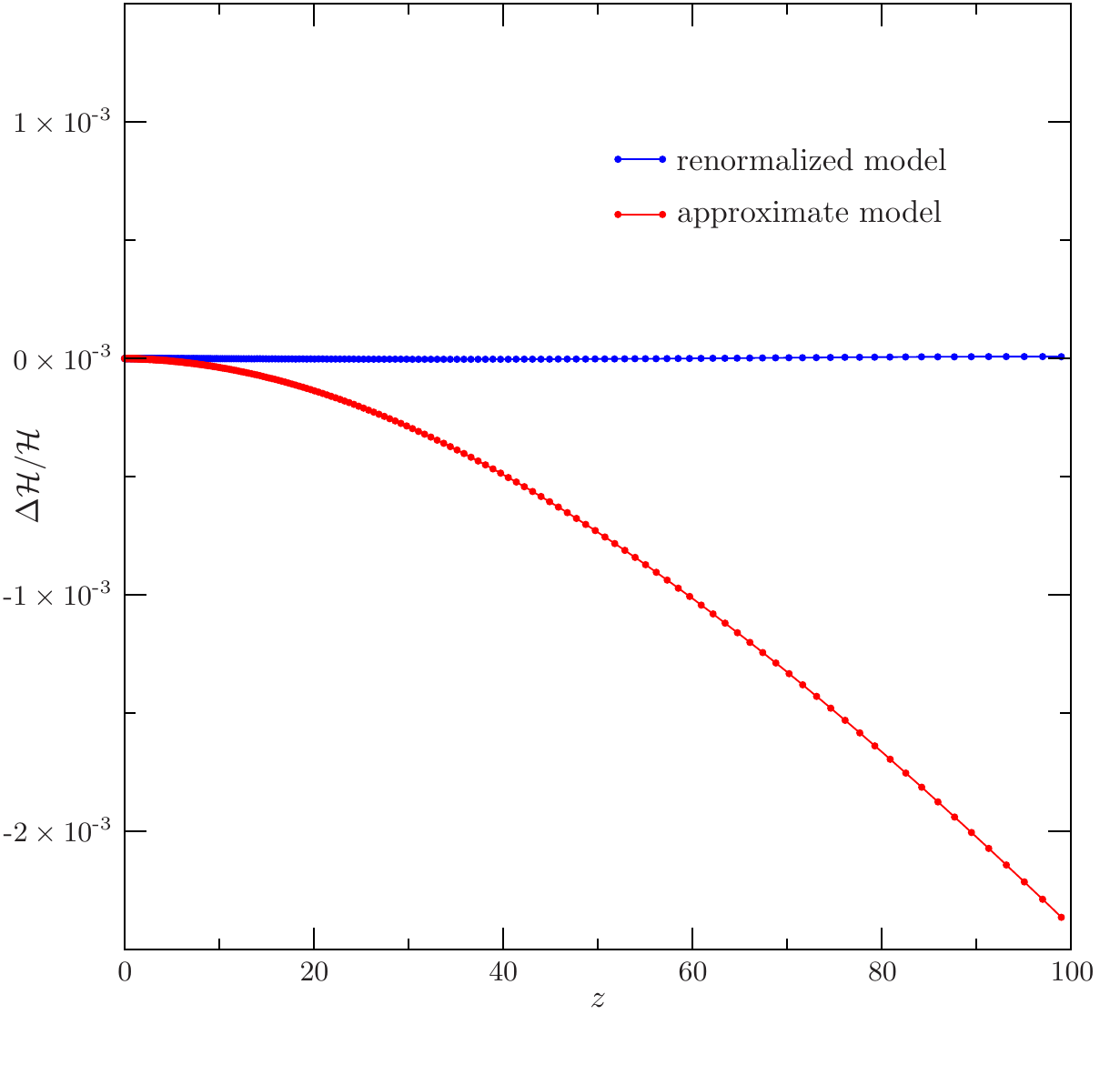}}
\caption{\label{fig:backreaction} We show the relative difference $\Delta \cH / \cH$ with respect to an ``exact'' FLRW reference model
for the numerically computed Hubble function of an approximate model (red) which neglects the relativistic kinetic energy of the
neutrino species. Using this approximate model in our N-body scheme results in a small ($\sim 10^{-4}$) but non-vanishing homogeneous mode
$\bar{\Phi} = \bar{\Psi}$ in the scalar metric perturbations. This homogeneous mode can be interpreted in terms of a renormalization of
the background model (see text), and we also show the $\Delta \cH / \cH$ for this renormalized model (blue), obtained numerically with our
N-body code. The renormalization procedure completely accounts for the inaccuracy of the approximate background model.
The numerical simulation was carried out with a lattice of $512^3$ points and a comoving box of $\sim 1$ Gpc/$h$. The remaining
parameters and the cosmological model are explained in Section \ref{sec:neutrino}.}
\end{figure}

To summarize, the fact that we have adopted an approximate background model gives rise to a homogeneous mode $\bar{\Phi}$ which
renormalizes the background. Figure \ref{fig:backreaction} shows the relative change of the Hubble function for the approximate and
the renormalized model, both with respect to the model where the neutrino energy densities are computed by integrating over the
unperturbed distribution functions, i.e.\ the background model which is considered the usual one in linear perturbation theory.
More specifically, $\Delta \cH \doteq \cH_\text{model} - \cH_\text{exact}$, where $\cH_\text{exact}$ is the conformal Hubble function obtained by solving
the FLRW reference model exactly, including the relativistic energy densities of the neutrino species.

As one can see from Figure \ref{fig:backreaction}, the approximate model deviates noticeably from the reference model at high redshift since
it misses the kinetic energy of the neutrinos which becomes gradually more important as one moves backwards in time. The renormalized model,
on the other hand, coincides almost perfectly with the reference model. We also checked that the homogeneous mode $\bar{\Phi}$, and
therefore the renormalization, goes down to nearly zero as soon as we adopt the more accurate background model $\cH_\text{exact}$.
Some tiny differences are expected due to the fact that beyond linear order there are
some terms which do not average to zero in eq.~(\ref{eq:E00}) but which are not taken into account in the reference model. These are, in
particular, the kinetic energy and binding energy of CDM perturbations \cite{Adamek:2014gva}. However, in our example, these effects lead to a much smaller renormalization of the background than the neglect of the neutrino kinetic energy.

What we have demonstrated with this exercise is the following. The precise choice of background model is indeed not important in our framework.
Any inaccuracy in the model is consistently accounted for when solving for the perturbations. In particular, the homogeneous mode in the
scalar metric perturbations renormalizes the background model towards a unique solution which could be called the ``resummed'' background model 
from the point of view of our gauge choice. One should keep in mind, however, that we work in a weak field expansion, and therefore this
renormalization scheme only works as long as the difference between the adopted model and the ``resummed'' one remains small. To make this
statement more precise, we require $|\bar{\Phi}| \ll 1$ at all times. In practice one can use the numerical value of $\bar{\Phi}$ as a
diagnostic to determine whether the true geometry is adequately described by the adopted background model.

It should also be stressed that the traditional Newtonian approach neglects any possible renormalization of the background by
construction. While it is possible to estimate the size of the error (see again \cite{Adamek:2014gva}), it is not straightforward
to correct for it within the course of the N-body evolution. The main reason is that Newtonian quantities, e.g.\ the density, are
computed in an \textit{unperturbed} geometry.

\subsection{Neutrino cosmology}
\label{sec:neutrino}

In this section we want to apply our relativistic N-body scheme to a cosmology with massive neutrinos. This is especially important as the effects of neutrinos on cosmological large scale clustering is one of the most promising paths to measure the presently
unknown absolute mass scale of neutrinos and it is one of the main goals of several large scale surveys which are presently planned~\cite{Font-Ribera:2013rwa,Amendola:2012ys,Laureijs:2011gra,Abate:2012za}.
The free streaming scale of neutrinos, $\lambda_\nu$, is given (very roughly) by the particle horizon at the time neutrinos become non-relativistic, i.e.\ when $T_\nu = (4/11)^{1/3}T_{\gamma} \simeq m_\nu$. Neutrino density fluctuations with wave numbers larger than $k_\nu \doteq 2 \pi / \lambda_\nu$ are damped by the free streaming of neutrinos. 
As only neutrino fluctuations are damped, the amount of damping of the matter power spectrum depends on $\Omega_\nu/\Omega_m$, which is also determined by the neutrino mass. Details on neutrino cosmology can be found in the monograph on the subject~\cite{2013:neutrino}.

Here, we focus
on two important effects. Firstly, the presence of a free streaming relativistic species (massive or not) gives rise to anisotropic
stress (sometimes called \textit{shear}) which is an important source for $\chi = \Phi - \Psi$. In fact, as long as the neutrinos are relativistic, they are the dominant
contribution to $\chi$ on scales larger than the free streaming scale $\lambda_\nu$. 
This is a relativistic effect which is not present in Newtonian N-body codes which only determine a Newtonian approximation to
$\Phi$ and set $\chi=0$.
The second effect comes from the fact that, as mentioned above, free streaming washes out density perturbations in the neutrino distribution. Massive neutrinos, which are non-relativistic at low redshift, form a component of the
total matter power spectrum which is smoother than that of CDM. Therefore, at fixed total matter density (including massive
neutrinos), a larger neutrino mass leads to a lower total matter power at small scales. This
 only affects scales below the free streaming scale  $\lambda_\nu$. On scales much larger than that, the free streaming has little effect
on the power spectrum.

The consistent treatment of neutrinos within N-body simulations constitutes a difficult problem which is the domain of an entire field of
research, see~\cite{2013:neutrino} for an overview. The smoothness of the neutrino density field suggests that a perturbative calculation could lead quite far. For instance, one can use the linear solution of the neutrino perturbations computed with a Boltzmann code and add their contribution to the density field in Fourier space, see e.g.\ \cite{Brandbyge:2008js}. This method can be improved to take into account
the potentials of nonlinearly evolved DM \cite{AliHaimoud:2012vj}, however, it remains difficult to account for the phase information in the neutrino perturbations. In this perturbative formalism the neutrinos are treated as an independent, linear  Gaussian random field and it fails to capture the full picture in the deeply nonlinear regime as the density contrast of neutrinos becomes larger than unity even for quite moderate neutrino masses.

Treating the neutrinos as N-body particles (see e.g.\ \cite{VillaescusaNavarro:2012ag,Inman:2015pfa}) seems conceptually straightforward and it should contain all the relevant
physics. The main difficulty here is the fact that the initial phase space distribution is very broad: it is an extremely
relativistic Fermi-Dirac distribution. This is true even in the non-relativistic regime at low redshift since neutrinos have decoupled while still relativistic and they cannot return to thermal equilibrium later due to their weak coupling. Sampling this distribution
with N-body particles is very inefficient and shot noise becomes a serious concern (this problem is absent within the perturbative approaches mentioned earlier). We 
nevertheless follow this approach and try to address the problem of shot noise as good as we can.

\subsubsection{Shot noise}

We  take two measures to mitigate shot noise. Firstly, we  ensure that the power spectrum of the perturbations in the total energy
density is not  significantly affected by shot noise on any scale resolved by the simulation. This is possible even in cases where the
neutrino density power spectrum is dominated by shot noise, as is the case on small scales and at high redshift when neutrinos are still relativistic. However, the neutrinos are only a small contribution to the total energy density, and it turns out that we can reduce their
shot noise to a level well below the total matter perturbation power.

\begin{figure}[t]
 \centerline{\includegraphics[width=0.62\textwidth]{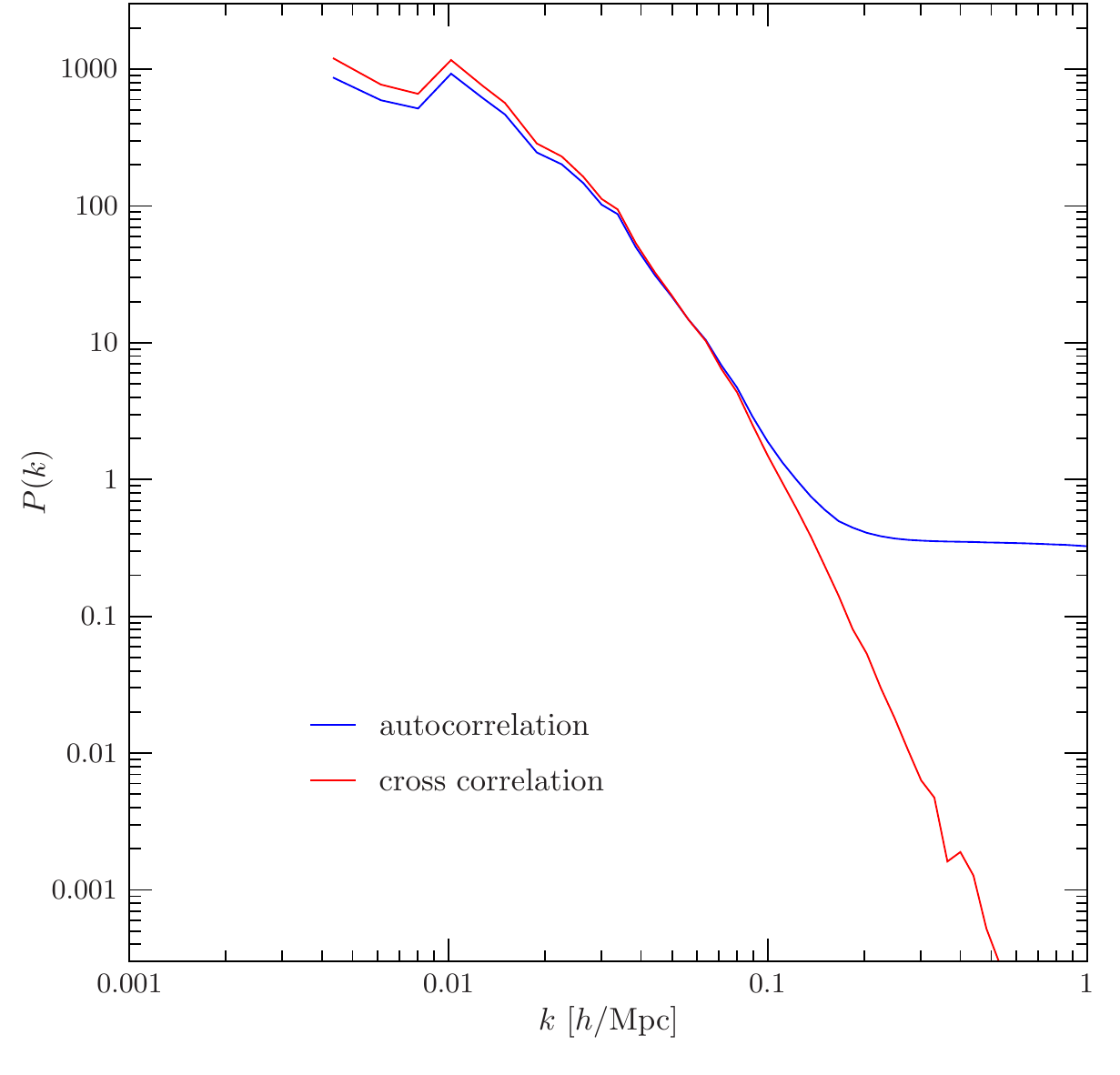}}
\caption{\label{fig:autovscross} We compare two numerical estimators for the neutrino power spectrum (taken at redshift $z = 3$): the blue curve shows the usual estimator using the autocorrelation of the density field, while the red curve is obtained by splitting the N-body ensemble into two
subensembles and computing the density cross correlation. For this particular simulation (which had $2\times2048^3$ neutrino particles
of the given flavor) the shot noise contribution completely dominates the autocorrelation at scales $k \gtrsim 0.2 h/$Mpc, while the cross correlation estimator is completely free of this type of contamination.}
\end{figure}

When computing the power spectrum of the neutrino density, and also the power spectra of any other quantity which linearly depends on the
neutrino stress-energy, such as $\chi$, we can use a trick to suppress the shot noise contribution \cite{Inman:2015pfa}: we simply split the neutrino
particle ensemble arbitrarily into two subensembles of equal particle number and then compute the cross-correlation power spectra between
the two subensembles. Since both subensembles sample the same distribution, the cross-correlation is a good proxy for the
autocorrelation. However, since the shot noise is uncorrelated between the subensembles, it partially cancels out, the degree of
cancellation depending on the number of modes contributing to each bin. Schematically,
denoting the full numerical density fluctuation by $\tilde\delta$, the physical density contrast by $\delta$ and the shot noise by $\eta$,
\begin{eqnarray}
 \tilde{\delta} = \delta + \eta &\quad\rightarrow\quad& \langle \tilde{\delta}\tilde{\delta} \rangle = \langle \delta \delta \rangle + 2 \cancel{\langle \delta \eta \rangle} + \langle \eta \eta \rangle \, ,\nonumber\\
 &\quad\rightarrow\quad& \langle \tilde{\delta}_1 \tilde{\delta}_2 \rangle = \langle \delta_1 \delta_2 \rangle + \cancel{\langle \delta_1 \eta_2 \rangle} + \cancel{\langle \delta_2 \eta_1 \rangle} + \cancel{\langle \eta_1 \eta_2 \rangle} \, .
\end{eqnarray}
 The split into two sub-ensembles gives two numerical amplitudes $\tilde{\delta}_1$, $\tilde{\delta}_2$, whose
shot noise contributions are entirely uncorrelated. Figure~\ref{fig:autovscross} illustrates how effective this trick works in practice.

\subsubsection{Simulation parameters and initial conditions}

As example we choose a cosmology with three massive neutrino species, with mass eigenstates of $0.06$ eV, $0.06$ eV, and $0.08$ eV, respectively. This corresponds roughly to a normal hierarchy where we neglect the smaller of the two experimentally known mass-square
differences. The remaining cosmological parameters are chosen compatible with current observations~\cite{Ade:2015xua}. In particular, we set $\Omega_c = 0.259$,
$\Omega_b = 0.0483$, $n_s = 0.9619$, $A_s = 2.215 \times 10^{-9}$ (at $0.05$ Mpc$^{-1}$), and $h = 0.67556$. The value of $h$ is only used
to convert the physical neutrino density to a dimensionless density parameter -- by choice of appropriate units all other computations
are independent of $h$.

Since our code does not include a treatment of baryonic physics we treat CDM and baryons as a single species,
ignoring the somewhat different clustering properties which are mostly important at small scales. The baryon acoustic oscillations
in the linear power spectrum which are generated in the coupled baryon-photon fluid prior to recombination are contained in our
initial conditions. However, late time baryonic effects in non-linear structure formation are not included. Photons are only taken into account at
the background level, i.e.\ we neglect the perturbations in the radiation field.
In the present treatment we also neglect primordial gravitational waves. These remain small contributions to the curvature at all times
and can be calculated to sufficient accuracy within linear perturbation theory. Since they are uncorrelated with scalar perturbations,
where it does not vanish, their effect can be simply added to the statistical quantities computed here, like e.g.\ the tensor perturbation spectrum.

\begin{figure}[t]
 \centerline{\includegraphics[width=\textwidth]{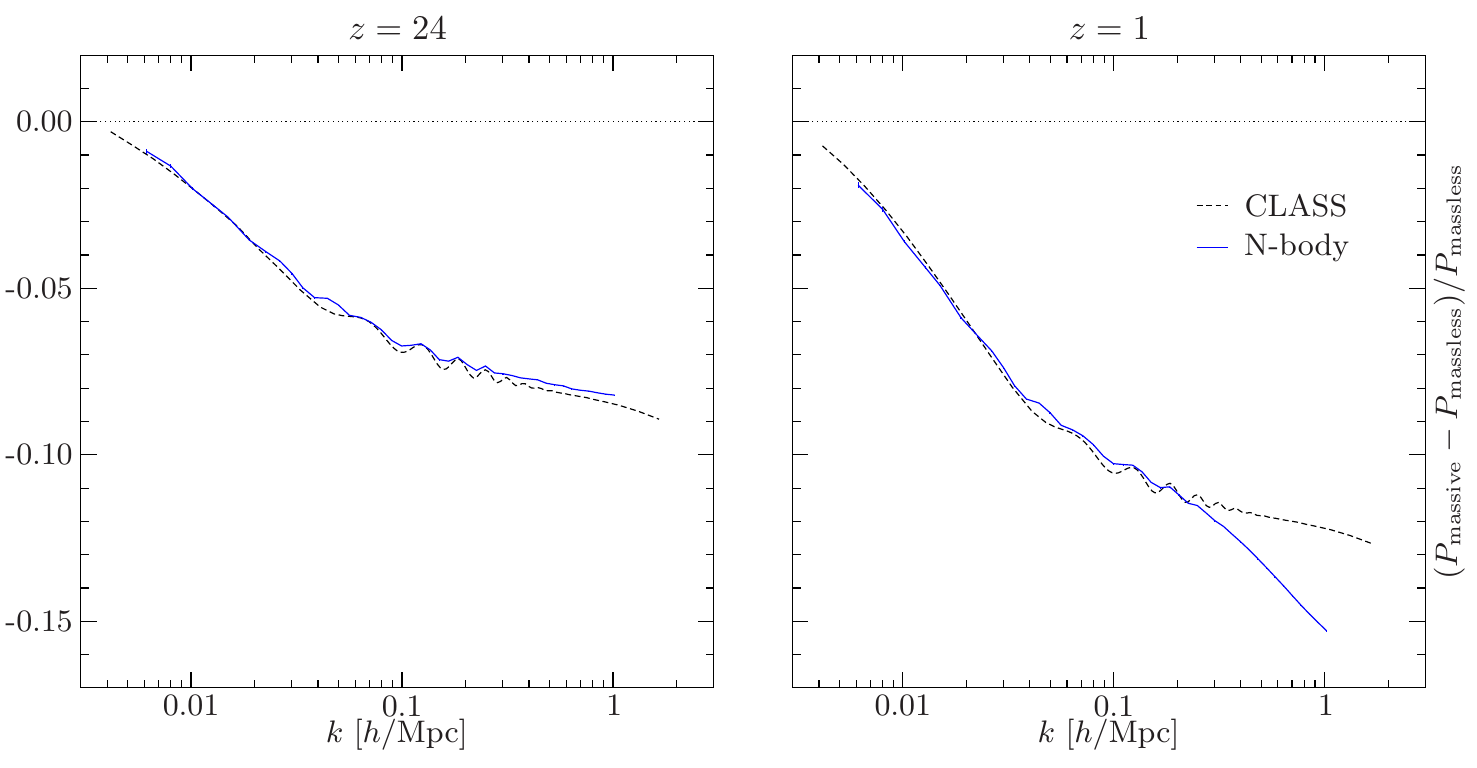}}
\caption{\label{fig:nudP}At two different redshifts we show the relative difference of the power spectra of $\Phi$ for a simulation
with massive neutrinos with respect to the massless case. In order to suppress cosmic variance, both simulations were initialized using
the same realization of a Gaussian random field. The dashed curve indicates the linear prediction obtained with the Boltzmann code \textit{CLASS}.}
\end{figure}

\begin{figure}[t]
 \centerline{\includegraphics[width=\textwidth]{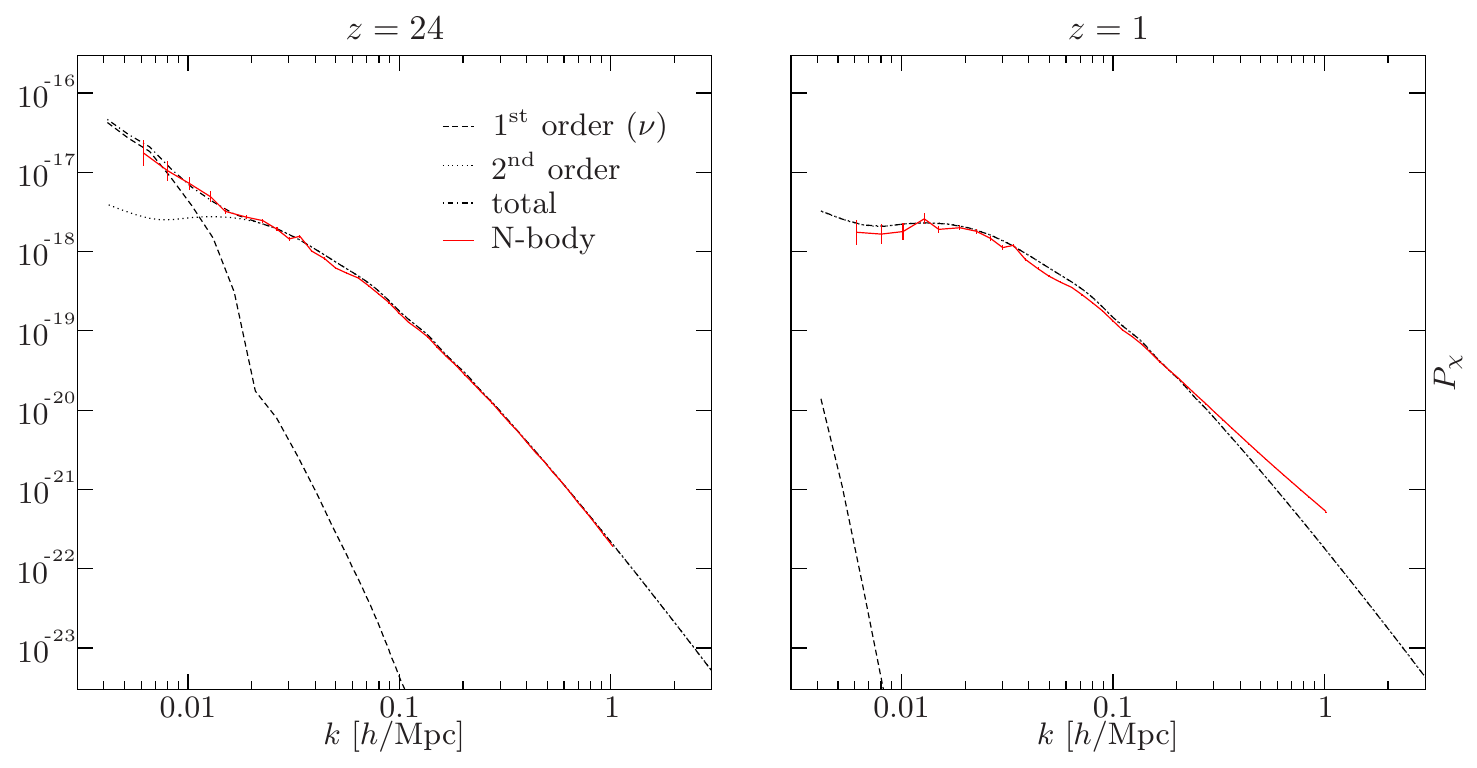}}
\caption{\label{fig:nuchi} We show the numerical power spectra (red) of $\chi = \Phi - \Psi$ at two different redshifts, together with
some perturbative calculations. The dashed curve indicates the contribution of the neutrino shear stress as predicted by linear theory
(the calculation was done with \textit{CLASS}), while the dotted curve is the contribution at second order which includes CDM and geometry.
The total (first and second order) is plotted as dot-dashed curve. At redshift $z = 24$ the numerical result agrees well with the perturbative
prediction, including the significant enhancement at large scales due to the neutrino shear stress. At $z = 1$ the N-body simulation
has additional power at small scales which is due to the fact that nonperturbative effects become relevant.}
\end{figure}

To set up initial conditions we follow a procedure proposed in \cite{Ma:1993xs}. Using a Boltzmann code \cite{Blas:2011rf}, we compute
the linear solutions (as a function of time) of $\sim 100$ perturbation modes, covering the entire range of scales represented in our
simulation. Next, we initialize the particle ensemble at redshift $z = 500\,000$, at which time the comoving particle horizon is just below our
lattice resolution and all modes in the simulation are superhorizon. At this time we may assume that the perturbations are adiabatic and
Gaussian. With this assumption, we draw a realization for $\Phi$ which also determines the initial particle positions, as well as the initial
values for $\Psi$. The amplitudes are determined from the $\sim 100$ linear mode functions using cubic spline interpolation in Fourier space. A random
momentum is added to the neutrino particles according to the initial Fermi-Dirac distribution. Details are given in Appendix~\ref{app:ICs}.

The phase space of the N-body ensemble is then evolved to the initial redshift of the N-body simulation, which was chosen as $z = 99$.
This is done by evolving each perturbation mode of the potentials $\Phi$ and $\Psi$ according to the linear solution, and then solving
the geodesic equation for the particles in these potentials. While this is a computationally expensive\footnote{About $11.5 \%$ of the computational time was spent pre-evolving the particles from $z = 500\,000$ to $z = 99$. The remaining $88.5 \%$ were attributed to the fully nonlinear N-body simulation from $z = 99$ to $z = 0$.} way to set up the initial particle phase space, it guarantees that all moments of
the distribution function and all relative phases are set up correctly, at least in principle. One may argue that much of this information
is swamped by shot noise, but this depends on the number of particles in the simulation. Our largest simulations had a lattice of
$2048^3$ cells and a total of $43$ billion neutrino particles and $4.3$ billion CDM particles. The box size for this simulation was chosen as
$1448$ Mpc/$h$, giving a resolution of $\sim 0.7$ Mpc/$h$. The integration used $1083$ time steps: $271$ steps for pre-evolving the particle
phase space in the linear potentials and $812$ steps for the nonlinear evolution. The entire simulation used some $40\,000$ CPU hours on
the Cray XC30 supercomputer \textit{Piz Daint}.

In addition to a simulation with massive neutrinos we also do a simulation for the massless case. In this case the neutrinos are simply
treated as part of the radiation field, with the standard value of $N_\text{eff} = 3.046$. The density of CDM is increased to
$\Omega_c = 0.264$ in order to have the same total matter density in both simulations (at $z = 0$). All other simulation parameters are
kept unchanged, in order to minimize systematics.

\subsubsection{Numerical results}

Figure \ref{fig:nudP} shows the relative difference in the perturbation power spectrum of $\Phi$ between the case of massive
and the case of massless neutrinos. The dashed line indicates the prediction in
linear perturbation theory, obtained with the Boltzmann code \textit{CLASS}. Our numerical results are in good agreement with the linear calculation
in the regime where we expect the latter to be valid. At late time and on small scales the linear theory breaks down, and we see that
the suppression of power in this regime is enhanced by nonlinear effects. This is quite intuitive: in the massless model the power is higher,
the modes become nonlinear quicker and hence have more time to increase their power through nonlinear growth.

As is well known, any free streaming relativistic species gives rise to anisotropic stress which sources $\chi = \Phi - \Psi$ even
at the linear level, see e.g.\ \cite{Ma:1995ey}. At second order also nonrelativistic species and even geometry contribute, the latter
by virtue of the nonlinear terms in the weak field expansion of the Einstein tensor, see eq.~(\ref{eq:ij}). Figure \ref{fig:nuchi} shows
the numerical power spectrum of $\chi$ at two different redshifts. We also indicate the contribution of the neutrino shear stress as computed
in linear perturbation theory, as well as the second order contribution from CDM and geometry. The radiation field also has some anisotropic
stress, but it is subdominant at these redshifts. We remind the reader that perturbations in the radiation field are neglected in
the simulation.

At $z = 24$ the neutrino shear stress is still the dominant contribution to $\chi$ at large scales $k \lesssim 0.01~h$/Mpc, while the
second order contributions from CDM and geometry dominate at smaller scales. Later, at $z = 1$, the neutrino shear stress has decayed due
to further redshifting of the neutrino distribution. At the smallest scales, $k \gtrsim 0.3~h$/Mpc, the power spectrum of $\chi$ gets
enhanced due to nonlinear evolution, an effect that is missed in the perturbative calculation which only goes to second order.

\begin{figure}[t]
 \centerline{\includegraphics[width=\textwidth]{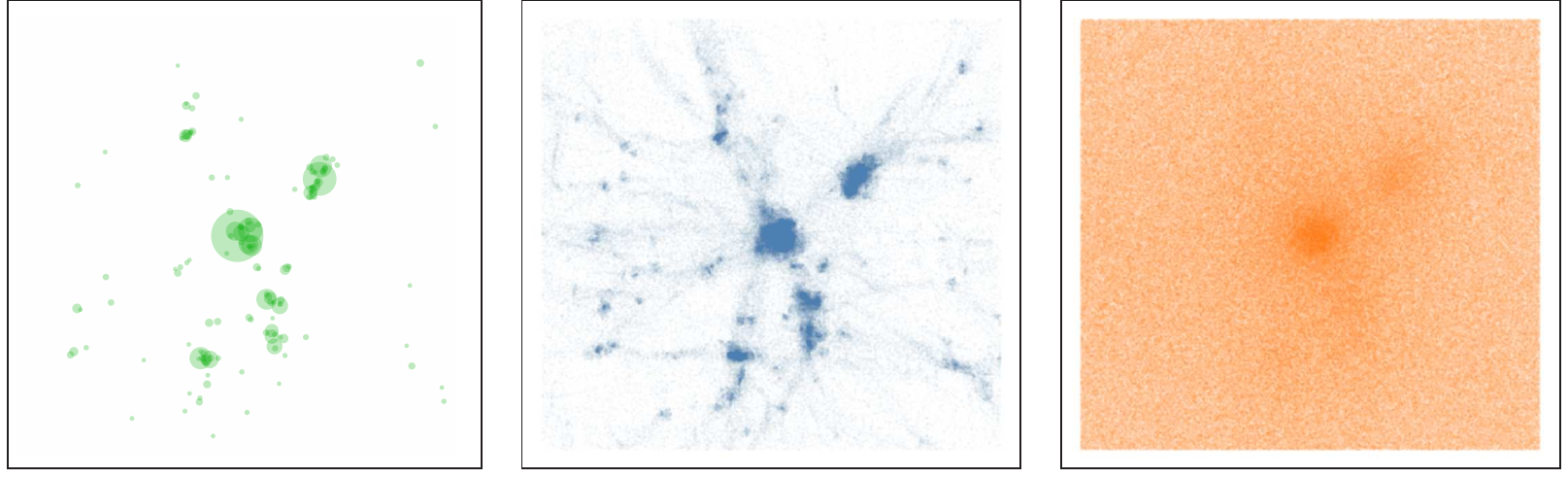}}
\caption{\label{fig:nuhalo}Zoom centered on a massive halo ($\sim 3\times 10^{15} M_\odot / h$) at $z = 0$.
The shown region is $50$ Mpc/$h$ across and has a depth of $20$ Mpc/$h$. The left panel shows the halos identified with the
\textit{Rockstar} halo finder using a conservative setting for the resolution. The disks are scaled to the virial radius as determined
with the halo finder algorithm. The center and right panels show the CDM and neutrino particles, respectively. The neutrino distribution
is very smooth due to the high thermal velocity dispersion. Only the deep potential wells of the most massive structures are able to
capture some neutrinos, giving rise to a diffuse neutrino halo around those objects.}
\end{figure}

Figure \ref{fig:nuhalo} shows a $50$ Mpc/$h$ region around a massive cluster at $z = 0$. As is evident from the right panel, the
neutrinos form a very smooth component of the matter distribution. Only the most massive objects are able to truly bind the neutrinos
within their potential wells. One should keep in mind that the mean thermal velocity for a $0.08$ eV neutrino is about $2000$ km/s
at $z = 0$, more than the escape velocity of most small scale structures.

\begin{figure}[t]
 \centerline{\includegraphics[width=0.62\textwidth]{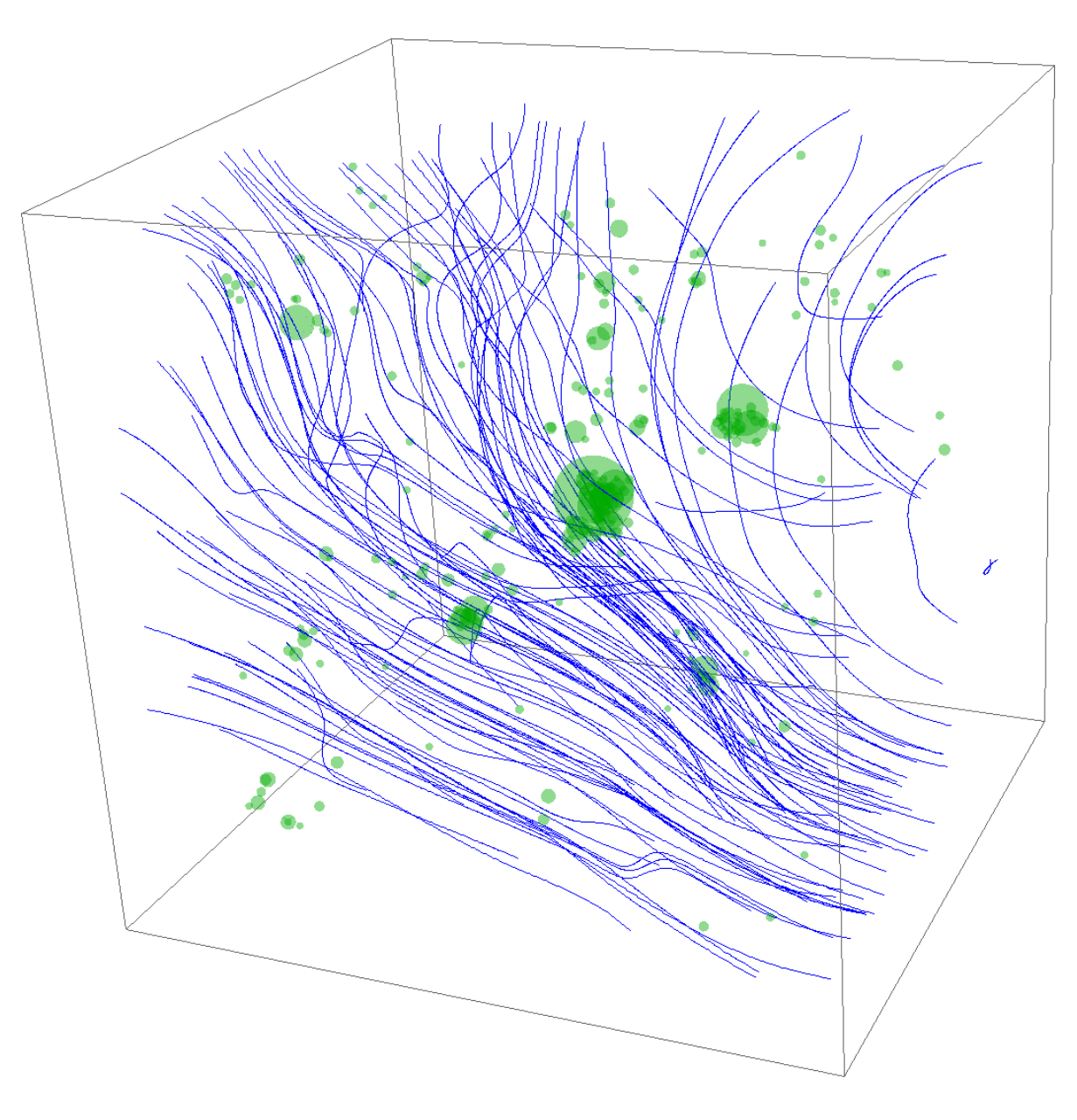}}
\caption{\label{fig:nuhalo3d} Three-dimensional rendering of the region shown in Figure \ref{fig:nuhalo}. CDM halos are
rendered as spheres scaled to the virial radius. The blue stream lines indicate the direction of $\nabla \times \mathbf{B}$
which plays the role analogous to the magnetic field in the gravitomagnetic Lorentz force equation (see text).}
\end{figure}

A three-dimensional rendering of the same region is shown in Figure \ref{fig:nuhalo3d} where we also include a stream plot of
the curl of the spin-1 metric component, $(\nabla\times \mathbf{B})^i \doteq \epsilon^{ijk} B_{k,j}$. Here we note that at leading
post-Newtonian order (i.e.\ at lowest order in inverse powers of the speed of light) the frame dragging force is given by a term
$m v \times (\nabla \times \mathbf{B})$ in the geodesic equation, see e.g.\ \cite{Adamek:2015eda}. Therefore the curl of $B_i$ plays
a role analogous to the one of the magnetic field in the Lorentz force (while the gradient of $\Psi$ is analogous to the electric field).
The magnitude of $\nabla \times \mathbf{B}$ in the particular volume shown in Figure \ref{fig:nuhalo3d} is of the order
of $10^{-23} h$/s in proper units. Due to the equivalence principle all nonrelativistic matter gyrates at this frequency. However, the corresponding period is five orders of magnitude larger than the present age of the Universe.

\subsubsection{Newtonian and relativistic particle trajectories}

In order to challenge the Newtonian approach used in traditional N-body codes we did a few smaller simulations using the Newtonian equations
of motion (\ref{eq:Newtonvel}), (\ref{eq:Newtonforce}) for the particle update
of the neutrinos. While this approximation introduces large errors on individual particle trajectories, the impact on expectation values
(such as the power spectrum) seems to be quite small. However, at early times the Newtonian propagation does not respect causality, as the velocity
of some particles becomes much larger than the speed of light. We therefore expect some errors to appear at the free streaming scale
$\lambda_\nu$.

If we define $\lambda_\nu$ precisely as the mean comoving distance covered by a neutrino particle between the
big bang ($z \rightarrow \infty$) and today ($z = 0$), one finds that for mass eigenstates of $0.06$ eV and $0.08$ eV the corresponding
relativistic free streaming scales are $1530$ Mpc/$h$ and $1317$ Mpc/$h$, respectively. Switching to the Newtonian approximation at
some initial redshift changes these numbers somewhat. For an initial redshift of $z=100$ the apparent free streaming length increases to
$1615$ Mpc/$h$ and $1361$ Mpc/$h$ for those neutrino masses, a respective change of $5.6 \%$ and $3.3 \%$. The change is larger
the smaller the neutrino mass and the earlier one chooses the initial redshift for the Newtonian evolution. It is quite possible
that percent errors on the free streaming length may propagate into the results of high-precision Newtonian simulations, but a
thorough investigation of this issue is left for future work.

We also note that the large distances travelled by the ``superluminal''
particles forces the Newtonian scheme to use a smaller time step. However, we think that all these issues can easily be mitigated by
making use of the relativistic equations of motion. For any Newtonian code simulating neutrinos we therefore propose to use what we
may call, somewhat imprecisely, ``Newtonian gravity with special relativity.'' By this we mean that we maintain the Poisson equation
(\ref{eq:Poisson}) which determines the Newtonian gravitational potential, but we change the equations of motion (\ref{eq:Newtonvel}), (\ref{eq:Newtonforce}) to
\begin{eqnarray}
 v_i &=& \frac{q_i}{\sqrt{q^2 + m^2 a^2}} \, ,\\
 \frac{\dd q_i}{\dd\tau} &=& -\frac{2 q^2 + m^2 a^2}{\sqrt{q^2 + m^2 a^2}} \left(\psi_N\right)_{,i} \, ,
\end{eqnarray}
which are approximations to the geodesic equations (\ref{eq:v_of_q}), (\ref{eq:geodesic}) that are however valid at all orders in the
velocity. Note the factor $2$ in front of the $q^2$ term which ensures the correct extremely relativistic limit: massless particles are deflected by $\Phi+\Psi$ which in the Newtonian limit is simply $2\psi_N$.

In this way it is guaranteed that the propagation of particles is causal, the free streaming length is computed accurately,
and even the deflection of relativistic ``test particles'' by
local potentials is treated correctly up to subleading weak field corrections. The above equations can easily be implemented in
Newtonian N-body codes, even those which use tree algorithms that are based on two-body forces.

\section{Summary}
\label{sec:summary}
In this paper we have presented in detail our new, publicly available relativistic N-body code. In a  previously published 
letter~\cite{Adamek:2015eda} we have shown the results for the relativistic degrees of freedom of the gravitational field, in particular 
frame dragging and gravitational waves. Here we explain in detail the theoretical underpinnings of the code, the approximation scheme 
used in our approach and the code structure. The details of the implementation of initial conditions, the particle to mesh projection and 
force interpolation as well as the Fourier space solver of the Einstein equations are presented in three appendices.
The source files and a manual are available on a public \textit{Git} repository:

\centerline{\texttt{\url{https://github.com/gevolution-code/gevolution-1.0.git}}}

After studying this paper we hope that people working in the field should be able not only to run the present version, but also to modify it for all interesting purposes like for studying models with dynamical dark energy or with modified gravity. The
code is distributed under a free software license and we encourage everyone to share their developments with the community.

In order to validate the code we perform two numerical tests. First we compare with an analytic solution by computing the metric of a single point mass. We find that it fares considerably better than the Newtonian solution and correctly includes also the terms in
the Schwarzschild metric which are of order $(r_S/r)^2$ and which are not present in a Newtonian approach.
Our implementation of the particle-mesh scheme is further validated using a Newtonian setting where we can compare to existing
N-body codes. Comparing to a simulation carried out with \textit{Gadget-2} we find that for a $\Lambda$CDM universe the power spectra of density perturbations agree within 1\% on scales larger than 10 times the Nyqvist frequency and within 2\% for scales larger than 5 times the Nyqvist frequency down to
redshift $z=0$.  Our code runs faster than \textit{Gadget-2}, but of course being a pure particle-mesh code based on field theory, it cannot
resolve scales smaller than the grid spacing. Gadget-2, on the other hand, follows the nonlinear evolution also on scales smaller 
than the Nyqvist frequency of its initial particle distribution.

In order to disentangle this resolution effect from other numerical differences we also compare to a simulation which was carried
out with \textit{RAMSES} at fixed spatial resolution, i.e.\ without using the adaptive mesh refinement capability of that Newtonian N-body code.
In this case the agreement remains within 1\% up to scales 3 times the Nyqvist frequency. As one approaches even closer to the
Nyqvist frequency one expects that differences in the discretization scheme become increasingly relevant.

Thanks to the highly scalable \lf\ library \cite{David:2015eya} \textit{gevolution} runs efficiently on the largest supercomputers with tens to hundreds of thousands of processor cores and supports very large grid sizes. Our relativistic code is therefore suitable especially on large to intermediate scales. On scales below $1$ Mpc/$h$ the modelling of structure formation is probably limited by our understanding of baryonic physics rather than by relativistic effects.

As opposed to Newtonian simulations where the Hubble function is an external input and dissociated from the simulation dynamics,
our relativistic approach jointly solves for background and perturbations in a self-consistent manner.
We demonstrate numerically that small changes in the background model are automatically  accounted for by the perturbations,
leading to a unique ``resummed'' geometry which is independent of the arbitrary split between background model and perturbations.
This renormalization procedure works as long as the assumed background model remains perturbatively close to the ``resummed'' one,
a requirement which can easily be diagnosed by monitoring the homogeneous mode of the scalar metric perturbation.

We finally show first results obtained with our code for a cosmology with massive neutrinos. Contrary to other procedures
used in the literature, which are based on Newtonian N-body codes, we treat the neutrinos fully relativistically and in a fully non-linear way from the beginning. We calculate $\chi = \Psi-\Phi$ which is dominated by free streaming neutrinos at early times and on large scales. Especially at late times we find no striking differences of our relativistic neutrino simulations compared to
the Newtonian ones, but a detailed comparison is non-trivial and  has to be presented in a future project.

\acknowledgments

We thank C.~Fidler, M.~Gosenca, S.~Hotchkiss, C.~Rampf, R.~Teyssier, T.~Tram and F. Villaescusa-Navarro for discussions. The numerical
simulations were carried out on \textit{Piz Daint} at the Swiss
National Supercomputing Centre (CSCS). This work was supported by the CSCS under project ID d45, and by the Swiss National Science Foundation.

\appendix
\section{Initial conditions}
\label{app:ICs}

The general procedure for setting up initial data for a simulation is as follows. First, one chooses an initial time at which perturbation theory is still valid. This allows to make the connection
with ``initial'' conditions set at the hot big bang using perturbative methods and known physics. These ``initial'' conditions are well constrained by observations of the cosmic microwave background.
In this appendix, we explain how initial data for non-interacting particle species (relativistic or not) can be constructed in \textit{linear} theory. The construction can be improved by including
higher orders in the perturbation series, but this is beyond the scope of this paper. We also do not provide a discussion of additional sources here, such as scalar fields. For every particular model with
new ingredients, one has to consider initial conditions appropriately. For the purpose of this work, we assume that it is sufficient to compute the initial perturbation amplitudes using
a linear Boltzmann code such as \textit{CAMB} or \textit{CLASS}. The output power spectra of such a code are the input we need in order to generate a \textit{realization} in terms
of metric perturbations and the particle phase space distribution.

It should be stressed that the generation of initial conditions for our relativistic scheme, even though it follows similar procedures, is not identical to the Newtonian case. This is partially owed to
the fact that gauge issues have to be considered and treated properly. Furthermore, our scheme is more complete in several respects. For instance, it is straightforward to take into account the fact
that the potentials have a decaying mode originating from the transition between radiation and matter domination. The presence of the radiation era has a noticeable effect at high redshifts, $z \gtrsim 100$,
which is notoriously difficult to treat within a Newtonian N-body scheme. Our new framework is designed in order to take care of precisely such relativistic aspects.\footnote{The anisotropic stress due to perturbations
in the radiation field, especially for relativistic neutrinos, causes a mismatch between $\Phi$ and $\Psi$ at the percent level even in
linear theory for $z \gtrsim 100$ and scales $\gtrsim 1$ Gpc. Since $\Phi$ and $\Psi$ both correspond to first-order gauge-invariant
quantities this effect is present in any gauge and can only be treated correctly if the two potentials are kept independent.
While this is the case in our framework, the current implementation still ignores perturbations in the radiation field. Therefore the
potentials inevitably still fail to match the ones computed with a linear Boltzmann code. This shortcoming could  be overcome in
the future, for instance by adding the linear perturbations of radiation to the N-body scheme.}

The particle positions are generated by the action of a linear displacement field $x^i \rightarrow x^i + \delta x^i(\mathbf{x})$ on a homogeneous ``template''. In the simplest case, the template
can be a regular arrangement of N-body particles similar to a crystal, but one could also use a template more similar to a glass, for instance. The displacement field corresponding to a particular
realization of the density perturbation can always\footnote{Since linear curl-type displacements do not change the density, this is true in linear theory even in the case where vector modes are present.
The presence of vector modes would only affect the initial velocities.} be written as a gradient, $\delta x^i = \delta^{ij} \xi_{,j}$, which can be worked out from the linearized version of eq.~(\ref{eq:00}),
\begin{equation}
\label{eq:Ds}
\Delta \Phi - 3 \cH \Phi' - 3 \cH^2 \Psi = \frac{3}{2} \cH^2 \sum_{\mathrm{species~}i} \Omega_{(i)} D_s^{(i)} \, ,
\end{equation}
where we introduce $D_s^{(i)} \doteq [\delta T^0_0 / \bar{T}^0_0 ]_{(i)}$, and $\Omega_{(i)}$ is the corresponding density parameter, $\Omega_{(i)} \doteq [\bar{T}^0_0]_{(i)} / [\bar{T}^0_0]_{\mathrm{total}}$.
We follow the notation of \cite{2008cmbg.book}.

To simplify the discussion, let us assume that one can separate all particle species into two classes at the initial redshift of the simulation, one where $q^2 \gg m^2 a^2$ is a good approximation and
one with $q^2 \ll m^2 a^2$. If a species is at the transition from being ultra-relativistic to non-relativistic, the initial redshift can be chosen differently in order to avoid this situation.
Inspecting eq.~(\ref{eq:T00}) it is easy to see\footnote{A simple way to see this is to go to the continuum limit where the sum over particles can be replaced by an integral over the initial particle
positions. The ``$\Delta \xi$'' arises as the Jacobian when changing the integral measure from $x^i$ -- where the distribution in position space is uniform -- to $x^i + \delta^{ij} \xi_{,j}$. See also
\cite{Adamek:2015hqa}.} that for nonrelativistic species, $q^2 \ll m^2 a^2$,
\begin{equation}
\label{eq:Dnonrel}
D_s^{(i)} = -\Delta \xi + 3 \Phi \, .
\end{equation}
For relativistic species there are various ways to implement a given density perturbation since the energy density depends on both, the number density and the particle momenta at leading order.
The simplest setting is the one where the phase space distribution is thermal, in which case the number density and momenta are uniquely given by a temperature. In this case one finds
\begin{equation}
\label{eq:Drel}
D_s^{(i)} = -\frac{4}{3} \Delta \xi + 4 \Phi \, ,
\end{equation}
and a perturbation $\delta T / T = D_s^{(i)} / 4$ is added to the temperature in order to determine the thermal momentum distribution. Incidentally, in the case of adiabatic initial conditions
where $D_s^{(i)} / [1+w_{(i)}] = D_s^{(j)} / [1+w_{(j)}$], the displacement field for all particle species (relativistic or not) is given by the same scalar field $\xi$. Here, the equation of state
$w_{(i)} = 0$ for nonrelativistic species, whereas $w_{(i)} = 1/3$ for particles that are ultra-relativistic. We assume adiabatic initial conditions from now on, such that eq.~(\ref{eq:Ds})
in Fourier space becomes
\begin{equation}
\label{eq:xi}
k^2 \Phi + 3 \cH \Phi' + 3 \cH^2 \Psi + \cH^2 \left[\frac{9}{2} \Omega_{\mathrm{non-rel.}} + 6 \Omega_{\mathrm{rel.}}\right] \Phi = -k^2 \cH^2 \left[\frac{3}{2} \Omega_{\mathrm{non-rel.}} + 2 \Omega_{\mathrm{rel.}}\right] \xi \, .
\end{equation}
If $\Phi$ and $\Psi$ are Gaussian random fields, then so is $\xi$, and its amplitude (rms value) can be obtained from the above equation as a function of the initial amplitudes of $\Phi$ and $\Psi$, their
deterministic mode evolution, and the known time-dependent background functions $\cH$, $\Omega_{\mathrm{non-rel.}}$ and $\Omega_{\mathrm{rel.}}$.

If adiabaticity is not a good approximation, for instance on sub-horizon scales for species which evolve differently after horizon entry  of the modes, the displacement field is not given by a relation as in eq.~(\ref{eq:xi}). In general, each species has its own displacement field, related to its density perturbation according to eq.~(\ref{eq:Dnonrel}) or eq.~(\ref{eq:Drel}). The displacement fields can therefore be constructed using the
respective transfer functions computed with a linear Boltzmann code. Note that these transfer functions are often computed in the comoving synchronous gauge whose density perturbation $D^{(i)}$ is related to $D_s^{(i)}$ in Fourier space as
\begin{equation}
 D^{(i)} = D_s^{(i)} - 3 \left(1 + w_{(i)}\right) \cH v^{(i)} \, ,
\end{equation}
where the scalar velocity potential $v^{(i)}$ will be introduced below.

This concludes the discussion of the initial particle displacement. To summarize, the initial condition generator implemented in \textit{gevolution} generates a Gaussian realization of $\xi$ according to
eq.~(\ref{eq:xi}) using the output of a linear Boltzmann code, and the displacement acts on a given homogeneous template. Mode by mode this also determines the initial amplitude of $\Phi$ and $\Psi$
corresponding to this particular realization. In the remaining part of this appendix we discuss how to set initial momenta.

In general, the momenta for an ensemble of particles are given by a phase space distribution function. Even in cases where this function
takes a simple form to begin with, its evolution can become very complicated which is the whole point of N-body simulations. For the purpose
of setting initial conditions we however assume that the phase space distribution is determined to a good approximation by its first
two moments. These are essentially given by the density, $-T^0_0$, and the \textit{momentum flux}, $T^0_i$, and we assume that the
anisotropic stress (the traceless part of $T^i_j$) can be neglected. This situation is often compared to a perfect fluid which is locally
described by its density and velocity fields.

We should stress that this so-called fluid approximation is a very restrictive assumption. While the stress-energy tensor can be constructed
for any phase space distribution function by means of eqs.~(\ref{eq:T00})--(\ref{eq:T0i}), the inverse is only possible in very special cases.
We discuss two distinct cases here. In the first case, which is the one relevant for non-relativistic species, the momentum flux of
the particle flow is much larger than the \textit{momentum dispersion}. This means that the phase space distribution function is essentially
a Dirac delta function in momentum space. In this case we do not care about the momentum dispersion, and the phase space distribution function
can be sampled with a single particle whose momentum is set by the momentum flux of the flow.

In the second case, relevant for ultra-relativistic species, the momentum flux is generally much smaller than the momentum dispersion. The
peculiar momenta are therefore dynamically important as they give rise, for instance, to a sound horizon and an effective pressure
which even affects the background. The distribution of momenta is very broad and can certainly not be approximated by a delta function.
For the purpose of setting initial conditions, we  assume that the distribution takes a simple parametric form. In particular, the momentum
flux is realized by displacing an isotropic distribution by an appropriate momentum vector in momentum space. An example would be a thermal
momentum distribution which is boosted into a frame to match the desired momentum flux.

In order to be more specific, let us discuss again the situation where the perturbations of stress-energy are generated by a mixture of
relativistic and non-relativistic species. We define a scalar potential $v$ for each species which
characterizes the curl-free part of the momentum flux as\footnote{Another frequently used quantity is the divergence of the velocity field, characterized by $\theta$ which is given by $\theta = -k^2 v$ in Fourier space, cf.\ \cite{Ma:1995ey}. Note also that \cite{2008cmbg.book} uses yet another quantity denoted as $V$ and related to $v$ as $V = -k v$. The motivation for this definition is that
$V$ is dimensionless. However, we prefer to work with $v$ since it is related to the velocity field (in configuration space) by a local operation.}
\begin{equation}
\label{eq:V}
 \frac{3}{2} \cH^2 \Omega_{(j)} \left(1+w_{(j)}\right) v_{,i}^{(j)} \doteq 4 \pi G a^2 \left[T^0_i\right]_{(j)} \, .
\end{equation}
In the linear regime the curl part can usually be neglected. Furthermore, for adiabatic initial conditions the potentials defined in
this way are identical for all species. We can therefore sum over all species and read off the potential from eq.~(\ref{eq:0i}),
\begin{equation}
\label{eq:Vad}
 \cH^2 \left[\frac{3}{2} \Omega_{\mathrm{non-rel.}} + 2 \Omega_{\mathrm{rel.}}\right] v = -\Phi' - \cH \Psi \, .
\end{equation}
By considering eqs.~(\ref{eq:T00}) and (\ref{eq:T0i}) this allows us to immediately set the initial momenta for non-relativistic
N-body particles as
\begin{equation}
\label{eq:ic_q}
 \frac{q_i}{m} = -\frac{2 a}{\cH^2} \frac{\Phi'_{,i} + \cH \Psi_{,i}}{3 \Omega_{\mathrm{non-rel.}} + 4 \Omega_{\mathrm{rel.}}} \, .
\end{equation}
Note that the momentum $q_i$ depends on the mass of the N-body particle which is a simulation parameter without any physical
significance. The code therefore internally works with the dimensionless quantity $q_i / m$.

The momenta of relativistic particles, however, are given by a large stochastic component with a simple parametric, isotropic distribution and
a comparatively small deterministic component which sets the mean momentum vector to
\begin{equation}
 \left\langle \frac{q_i}{m}\right\rangle = -\frac{8}{3 \cH^2} \frac{\Phi'_{,i} + \cH \Psi_{,i}}{3 \Omega_{\mathrm{non-rel.}} + 4 \Omega_{\mathrm{rel.}}} \left\langle\sqrt{\frac{q^2}{m^2}}\right\rangle \, ,
\end{equation}
where $\langle\cdot\rangle$ denotes a (mass-weighted) mean over the particle distribution at fixed position. The main difference to
eq.~(\ref{eq:ic_q}) is the fact that the momentum flux is realized in an average sense, and that
the individual momenta can be vastly different from this average. The last term on the right hand side is the mean energy
per particle (in the ultra-relativistic limit).

Adhering to the N-body concept, we model the phase space distribution by a large number of particles whose momenta are randomly drawn
from the distribution function. However, the number of particles is limited by computational resources and therefore the sampling of
phase space is generally  very sparse. In our simulation reported in Section~\ref{sec:neutrino}, we admit 5 neutrinos per lattice
site, each with momentum $q_{(n)} = \langle q\rangle + \delta q_{(n)}$, where $\delta q_{(n)}$ is randomly sampling the momentum
distribution and is typically much larger than $ \langle q\rangle $. For a broad distribution (like the extremely relativistic Fermi-Dirac distribution of cosmic neutrinos) shot noise becomes a serious issue. 
In the power spectra we can reduce the noise significantly by cross-correlating neutrinos from two sub-ensembles which sample the same distribution. Furthermore, the relativistic species give a \textit{subdominant} contribution to the energy
density, so that the imprint of the noise onto the gravitational potential is negligible.

Note again that eq.~(\ref{eq:Vad}) assumes adiabaticity; in situations where this assumption is dropped one can instead employ
eq.~(\ref{eq:V}) directly, using the transfer functions of $v^{(i)}$ as computed for each species separately. For non-relativistic species
this equation implies
\begin{equation}
 \frac{q_i}{m} = a v_{,i} \, ,
\end{equation}
whereas the ultra-relativistic limit reads
\begin{equation}
 \left\langle \frac{q_i}{m} \right\rangle = \frac{4}{3} \left\langle\sqrt{\frac{q^2}{m^2}}\right\rangle v_{,i} \, .
\end{equation}

\section{Particle-to-mesh projection and force interpolation}
\label{app:PM}

In any particle-mesh N-body scheme there are two different concepts for the representation of simulation data. One set of data represents
a phase space distribution. It is most conveniently discretized by sampling it with N-body particles. The particle ensemble
is stored as a list with six real-valued phase space coordinates per particle, the position and the momenta of the particles. Hence particles do not sit on the lattice sites. These coordinates are real-valued in order to allow for smooth
particle trajectories. Another set of data represents fields. Such data is most conveniently discretized by sampling the field values
on a discrete lattice which spans the three-dimensional simulation volume.

In order to couple the evolution of fields to the evolution of particles (and vice versa), one needs to provide methods which connect
the two different ways the data is represented. This is achieved by two types of operations which are dual to one another. The 
\textit{particle-to-mesh projection} converts some information contained in the particle ensemble into fields. In a Newtonian code one would,
for instance, compute a density field from a given particle distribution. In our relativistic framework the quantity of interest is the
stress-energy tensor of the particle ensemble. The dual operation is often called \textit{force interpolation} because the Newtonian force on each particle is determined by interpolation of the gravitational potential to the particle position. We  adhere to this terminology
even though interpolation is not only used for forces, but any time we need to determine the value of a field at the position of a particle.

A systematic way to construct these operations can be found in textbooks, e.g.\ \cite{HockneyEastwood}. For our purpose it is useful to
define two one-dimensional weight functions,
\begin{equation}
 {\sqcap}(b) \doteq \begin{cases}
                  1 & \text{if}~ -\frac{\dd \mathrm{x}}{2} < b < \frac{\dd \mathrm{x}}{2} \\
                  0 & \text{otherwise}
                 \end{cases} \, ,
\end{equation}
and
\begin{equation}
 {\wedge}(b) \doteq \begin{cases}
                      1+\frac{b}{\dd \mathrm{x}} & \text{if}~ -\dd \mathrm{x} < b < 0 \\
                      1-\frac{b}{\dd \mathrm{x}} & \text{if}~ 0 \leq b < \dd \mathrm{x} \\
                      0 & \text{otherwise}
                     \end{cases} \, ,
\end{equation}
where $\dd \mathrm{x}$ denotes the lattice unit, i.e.\ the coordinate distance between neighboring lattice points. The utility of these
weight functions becomes evident when considering the respective constructions of a (one-dimensional) density field,
\begin{eqnarray}
 \rho_\mathrm{NGP}(x_\mathbf{i}) &\doteq& \sum_{n} m_{(n)} {\sqcap}(x_\mathbf{i}-x_{(n)}) \, ,\\
 \rho_\mathrm{CIC}(x_\mathbf{i}) &\doteq& \sum_{n} m_{(n)} {\wedge}(x_\mathbf{i}-x_{(n)}) \, ,
\end{eqnarray}
where $m_{(n)}$ and $x_{(n)}$ are the mass and position of the $n$th particle, and $\mathbf{i}$ is a discrete index labelling the lattice points. The subscript \textit{NGP} refers to \textit{nearest grid point} assignment since using the weight
function $\sqcap$ corresponds to assigning all the mass of a particle to the nearest lattice point. This is often considered the zeroth-order approximation to the density field.
\textit{CIC} stands for \textit{cloud-in-cell}, and the weight function $\wedge$ corresponds to distributing the mass of a particle between
the two neighboring lattice points in such a way that the mass assignment changes linearly with the particle position. Higher order assignment schemes can be constructed but we do not discuss them here.

The weight functions also naturally define corresponding interpolation operations,
\begin{eqnarray}
 \psi_\mathrm{NGP}(x_{(n)}) &\doteq& \sum_{\mathbf{i}} \psi(x_\mathbf{i}) {\sqcap}(x_\mathbf{i}-x_{(n)}) \, ,\\
 \psi_\mathrm{CIC}(x_{(n)}) &\doteq& \sum_{\mathbf{i}} \psi(x_\mathbf{i}) {\wedge}(x_\mathbf{i}-x_{(n)}) \, .
\end{eqnarray}
Note that the sum is now performed over the lattice points.

\begin{figure}[t]
 \centerline{\includegraphics[width=0.62\textwidth]{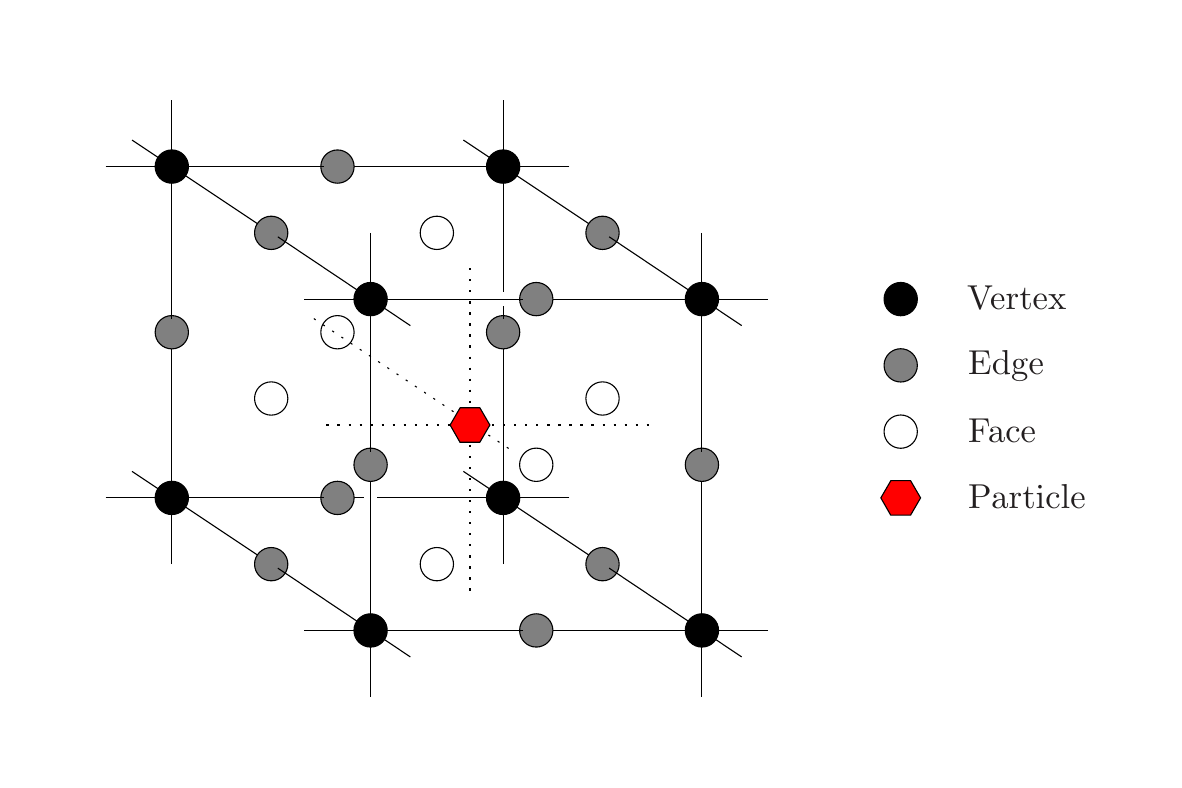}}
\caption{\label{fig:PM}Sketch of a lattice cell indicating the locations associated with various lattice quantities. Scalar fields, as well as the diagonal components of tensor fields, are associated with the vertices of the primary lattice. The components of a vector
field, including also gradient fields of scalars, are associated with the edges of the primary lattice. This association follows naturally
if one considers the discrete gradient operator implemented by taking the finite difference of adjacent sites, i.e.\ the one-sided two-point gradient. Finally, the off-diagonal components of tensor fields are associated with the faces of the primary lattice. The projection of particle
data onto the lattice sites is constructed using various combinations of piecewise constant or piecewise linear functions which take the particle coordinates within the lattice cell (dotted lines) as arguments.}
\end{figure}

\pagebreak
Another aspect of the lattice discretization concerns the definition of discrete derivative operators. Here we  always employ
the one-sided two-point gradient, which in one dimension is defined as
\begin{equation}
 (\nabla \psi)(x_\mathbf{i+\frac{1}{2}}) \doteq \frac{\psi(x_\mathbf{i+1}) - \psi(x_\mathbf{i})}{\dd\mathrm{x}} \, .
\end{equation}
Note that this type of gradient would naturally be associated with a lattice which is shifted by half a lattice unit. The three dimensional
generalization of this gradient operator has the three components of the gradient being associated with three different shifted lattices,
as the shift always applies in the direction in which the derivative is taken.

We apply this prescription to all vector-valued fields. Figure \ref{fig:PM} shows a sketch of a lattice cell indicating the various locations
to which field values can be associated. All scalar fields are associated with the vertices of the primary lattice. The components of a vector field are associated to the edges connecting the vertices, each possible orientation of the edges corresponding to one component of
the vector. Note that this convention ensures that the divergence of a vector field -- which is a scalar -- is again  associated with the vertices if one uses our discrete choice of derivative operator. Finally, the off-diagonal elements of a tensor field are associated to
the faces of the lattice cell, while the diagonal components are represented once more at the vertices. This ensures again the compatibility with our previous conventions when constructing vectors or scalars by applying discrete derivative operations on a tensor field (and vice versa).
Note also that the curl acting on a vector field associated with the edges gives a result which is naturally associated with the faces,
and vice versa. This is due to the fact that the curl of a vector is truly an anti-symmetric tensor, $\dd_iv_j-\dd_jv_i$, which only in 3 dimensions can be represented as a (pseudo-)vector. A double curl does not change the association with lattice sites, and neither does the gradient acting on a divergence.

Let us now turn back to the problem of particle-mesh projection and force interpolation. In order to construct a scalar quantity
(like the rest-mass density in a Newtonian setting or, for instance, $T^0_0$ in our framework) from an N-body ensemble we 
employ the three-dimensional version of the CIC operation. This means that we evaluate eq.~(\ref{eq:T00}) at the lattice points
by replacing
\begin{equation}
\label{eq:scalarPM}
 \delta^{(3)}(\xlat{}{}{}-\mathbf{x}_{(n)}) \rightarrow \prod_{i=1}^3 {\wedge}(x^i_{\mathbf{i},\mathbf{j},\mathbf{k}}-x^i_{(n)}) 
\end{equation}
in that equation. Here and in the following, $\mathbf{i}$, $\mathbf{j}$, $\mathbf{k}$ are the discrete indices labelling the vertices of the
three dimensional lattice. The geometric corrections brought by the $\Phi$-dependent terms in eq.~(\ref{eq:T00}) are incorporated by evaluating $\Phi$ directly
at the lattice points\footnote{\label{fn:PM}This choice is arbitrary and other choices lead to different discretization errors. As the
geometric corrections themselves are small we are not so much interested in the discretization error on them, and therefore make
a choice based on convenience of implementation.}. In Newtonian theory these corrections are not present and $T^0_0$ essentially reduces to the CIC rest-mass density.

One could, of course, proceed with the same prescription also for the construction of vector or tensor fields. However, we adhere
to a somewhat different construction principle. We require that the projection and interpolation operations for each particle
only depend on the lattice quantities which are associated with the boundaries of the lattice cell containing the particle.
These are the quantities associated with the eight vertices, the twelve edges and the six faces belonging to that cell (see Figure \ref{fig:PM}).

The vector projector used to construct $T^0_i$ is obtained by replacing
\begin{equation}
\label{eq:vectorPM}
 \delta^{(3)}(\xlat{+\frac{1}{2}}{}{}-\mathbf{x}_{(n)}) \rightarrow {\sqcap}(x^1_{\mathbf{i+\frac{1}{2}},\mathbf{j},\mathbf{k}}-x^1_{(n)}) \prod_{i=2}^3 {\wedge}(x^i_{\mathbf{i+\frac{1}{2}},\mathbf{j},\mathbf{k}}-x^i_{(n)})
\end{equation}
in eq.~(\ref{eq:T0i}) for the component $T^0_1$ which is associated with the lattice edge sites $\xlat{+\frac{1}{2}}{}{}$, and analogous
replacements for the other two vector components. This assignment scheme is a mixture between the NGP and CIC method. Observing that
\begin{equation}
\label{eq:PMrecursion}
 \frac{d {\wedge}}{d b} = \frac{{\sqcap}(b + \frac{\dd\mathrm{x}}{2}) - {\sqcap}(b - \frac{\dd\mathrm{x}}{2})}{\dd\mathrm{x}}
\end{equation}
reveals a special property of this assignment scheme. Namely, if the rest-mass density $\rho$ is constructed on the lattice using
CIC assignment, and the momentum field $\rho v^i$ is constructed using the vector projection above, the continuity equation
\begin{equation}
 \left[a^3 \rho\right]' + a^3 \left[\rho v^i\right]_{,i} = 0
\end{equation}
is satisfied \textit{identically} on the lattice with our definition of the discrete divergence operator. The general relativistic counterpart
to this equation is the time component of the covariant conservation equation,
\begin{equation}
\label{eq:energyconstraint}
 \nabla_\mu T^\mu_0 = \left[T^0_0\right]' + \left[T^i_0\right]_{,i} + \Gamma^i_{i0} T^0_0 + \Gamma^j_{ij} T^i_0 - \Gamma^i_{00} T^0_i - \Gamma^j_{i0} T^i_j = 0 \, .
\end{equation}
At zeroth order in the metric perturbations the only non-vanishing Christoffel symbols are $\Gamma^j_{i0} = \Gamma^j_{0i} = \delta^j_i \cH$, $\Gamma^0_{ij} = \delta_{ij} \cH$ and $\Gamma^0_{00} = \cH$. To satisfy eq.~(\ref{eq:energyconstraint}) at zeroth order exactly when using the
CIC assignment scheme (\ref{eq:scalarPM}) for $T^0_0$, one not only has to choose the assignment (\ref{eq:vectorPM}) for the vector
field $T^i_0$: interestingly, the CIC assignment scheme (\ref{eq:scalarPM}) has to be chosen also for the diagonal components of $T^i_j$.

What about the off-diagonal elements of $T^i_j$? Looking at the spatial part of the covariant conservation equation,
\begin{equation}
\label{eq:momentumconstraint}
 \nabla_\mu T^\mu_j = \left[T^0_j\right]' + \left[T^i_j\right]_{,i} + \Gamma^\mu_{\mu 0} T^0_j + \Gamma^\mu_{\mu i} T^i_j - \Gamma^0_{0j} T^0_0 - \Gamma^0_{ij} T^i_0 - \Gamma^i_{0j} T^0_i - \Gamma^k_{ij} T^i_k = 0 \, ,
\end{equation}
again at the zeroth order, one finds that the best choice seems to be the replacement
\begin{equation}
\label{eq:tensorPM}
 \delta^{(3)}(\xlat{+\frac{1}{2}}{+\frac{1}{2}}{}-\mathbf{x}_{(n)}) \rightarrow {\wedge}(x^3_{\mathbf{i+\frac{1}{2}},\mathbf{j+\frac{1}{2}},\mathbf{k}}-x^3_{(n)}) \prod_{i=1}^2 {\sqcap}(x^i_{\mathbf{i+\frac{1}{2}},\mathbf{j+\frac{1}{2}},\mathbf{k}}-x^i_{(n)})
\end{equation}
in eq.~(\ref{eq:Tij}) to compute the projection for $T^1_2$ at the lattice face $\xlat{+\frac{1}{2}}{+\frac{1}{2}}{}$, and analogous expressions for the other off-diagonal elements. This choice ensures that the contributions to eq.~(\ref{eq:momentumconstraint}) from the off-diagonal quantities cancel exactly with corresponding terms coming from $[T^0_j]'$. Unfortunately there remains a residual coming from the one diagonal term of $T^i_j$ which contributes to this equation. It seems difficult to satisfy the condition which makes this residual vanish, in particular if one wants to maintain also eq.~(\ref{eq:energyconstraint}). The same problem is already present in the Newtonian framework if one considers the Euler equation which is the nonrelativistic counterpart of eq.~(\ref{eq:momentumconstraint}). At the present stage this remains an artefact of
the lattice discretization that we have to bear with.

To summarize, with the choices of projection operations given above, the following coherent picture emerges. The contributions of each N-body
particle to the stress-energy tensor are projected onto lattice sites belonging to the cell in which the particle resides (see Figure \ref{fig:PM}). For components
associated with the vertices of the lattice, such as $T^0_0$ and the diagonal elements of $T^i_j$, the projection weights for the eight relevant vertices are piecewise trilinear functions of the three particle coordinates, corresponding to a three-dimensional CIC assignment. For components associated with edges
of the lattice, such as $T^i_0$ and $T^0_i$, the projection weights for the four relevant edges (per vector component) are given by piecewise bilinear functions of the two particle coordinates orthogonal to the edge direction, corresponding to a two-dimensional CIC assignment.
Along the coordinate of the edge direction the weights are piecewise constant as in the NGP assignment. Finally, for components associated with the faces of the lattice, such as the off-diagonal elements of $T^i_j$, the projection weights are piecewise linear functions of the particle coordinate orthogonal to the face plane, corresponding to a one-dimensional CIC assignment. Along the other two coordinates the weights are piecewise constant. Discretization errors are absent in eq.~(\ref{eq:energyconstraint}) and minimal (in some sense) in eq.~(\ref{eq:momentumconstraint}), at zeroth order in the metric perturbations\footnote{At first order in the metric
perturbations the discretization errors depend on the way one chooses to evaluate the first-order contributions to the Christoffel symbols as well as on various other choices (see also footnote \ref{fn:PM}). They are therefore not a property of the particle-to-mesh projection alone.}.

Our force interpolation is simply the dual of this prescription. For instance, in order to evaluate a vector field at a location inside
a lattice cell, the values specified on the edge sites of that cell are interpolated using the bilinear weights of the two-dimensional CIC assignment. In short, the interpolation amounts to an appropriately weighted average of the field values present at sites on the boundary of the cell.

\section{Fourier-space solvers for Einstein's equations}
\label{app:FFTsolvers}

In this appendix we explain the numerical algorithms implemented in our code for solving eqs.~(\ref{eq:00}) and (\ref{eq:ij}) in order to determine
the metric perturbations $\Psi$, $\Phi$, $B_i$ and $h_{ij}$ on the lattice. Let us begin with eq.~(\ref{eq:00}) which is a nonlinear parabolic (diffusion-type)
equation for $\Phi$. We  employ a first-order in time solution strategy, which allows us to effectively decouple the equation from the other set of Einstein's
equations at each update step. This is because we can choose to evaluate the $\Psi$-dependent term at the initial time where it is already known. In order to ensure
unconditional stability of the solution strategy, we have to use an implicit method (see \cite{NumericalRecipes} for more details). In the following, we  label
the time steps $\taulat{}$ by a discrete index $\mathbf{n}$ and the lattice points $\xlat{}{}{}$ again by discrete indices $\mathbf{i}$,
$\mathbf{j}$, $\mathbf{k}$. The finite difference between time steps are denoted as $\dd\tau \doteq \taulat{+1} - \taulat{}$, and the lattice unit is denoted as $\dd\mathrm{x}$ as before. Let us also introduce the shorthand $\lat{f}{n}{}{}{} \doteq f(\taulat{},\xlat{}{}{})$.
With these conventions, we can write a finite-difference version of eq.~(\ref{eq:00}) as
\begin{multline}
\label{eq:discrete00}
 \frac{\lat{\Phi}{n+1}{-1}{}{} + \lat{\Phi}{n+1}{+1}{}{} + \lat{\Phi}{n+1}{}{-1}{} + \lat{\Phi}{n+1}{}{+1}{} + \lat{\Phi}{n+1}{}{}{-1} + \lat{\Phi}{n+1}{}{}{+1} - 6 \lat{\Phi}{n+1}{}{}{}}{\dd\mathrm{x}^2} - 3\cH\frac{\lat{\Phi}{n+1}{}{}{}-\lat{\Phi}{n}{}{}{}}{\dd\tau} \\- 3\cH^2 (\lat{\Phi}{n}{}{}{}-\lat{\chi}{n}{}{}{})
 + \frac{3}{2} \frac{\left(\lat{\Phi}{n}{+1}{}{}-\lat{\Phi}{n}{-1}{}{}\right)^2 \! \!+\! \left(\lat{\Phi}{n}{}{+1}{}-\lat{\Phi}{n}{}{-1}{}\right)^2\! \! +\! \left(\lat{\Phi}{n}{}{}{+1}-\lat{\Phi}{n}{}{}{-1}\right)^2}{4\dd\mathrm{x}^2}\\ = -4 \pi G a^2 \left(1-4\lat{\Phi}{n}{}{}{}\right) \left[T^0_0(\taulat{+1}, \xlat{}{}{}) - \bar{T}^0_0(\taulat{+1})\right] \, ,
\end{multline}
where we have divided out a factor $(1+4\Phi)$ to move it to the right hand side. The implicit character of this finite-difference equation comes from the prescription to
evaluate the discrete Laplace operator at the final time $\taulat{+1}$. Had we chosen
to evaluate it at $\taulat{}$ instead, we would obtain an explicit equation which could be solved for $\lat{\Phi}{n+1}{}{}{}$ immediately. However, this strategy would impose
a condition on $\dd\tau$ in order to maintain stability. This so-called Courant-Friedrichs-Lewy condition requires that information may not travel farther than one lattice unit
in each update, i.e.\ $\dd\tau < \dd\mathrm{x}$. Such a requirement would render high-resolution simulations prohibitively expensive.

Once we have ensured stability, the choice of when to evaluate any of the remaining terms is not important in a first-order scheme. For convenience, we evaluate the stress-energy
tensor at $\taulat{+1}$ but we compute the geometric corrections using the values of the perturbations at $\taulat{}$. Also the remaining terms on the left hand side are evaluated
at $\taulat{}$. In a second-order in time integration scheme these choices would have to be modified, and all the evolution equations would have to be solved concurrently (for instance,
$\lat{\chi}{n+1}{}{}{}$ would occur in the above equation as well). This could be achieved using a predictor-corrector method, but we do not discuss this here since a first-order in time scheme
is sufficient for our purpose. Note also that the Poisson equation used to compute the Newtonian potential would effectively be solved using a zeroth-order (in time) scheme, as time derivatives
do not occur in this approximation at all.

In order to solve eq.~(\ref{eq:discrete00}), the code collects all the known terms, i.e.\ the stress-energy tensor and the terms built with known field values (those at $\taulat{}$),
and combines them to a single (nonperturbative) source field $\lat{s}{}{}{}{}$,
\begin{multline}
 \lat{s}{}{}{}{} \doteq 3\cH^2 \left(\lat{\Phi}{n}{}{}{} - \lat{\chi}{n}{}{}{}\right) - \frac{3}{8 \dd\mathrm{x}^2} \left(\lat{\Phi}{n}{+1}{}{}-\lat{\Phi}{n}{-1}{}{}\right)^2 - \frac{3}{8 \dd\mathrm{x}^2} \left(\lat{\Phi}{n}{}{+1}{}-\lat{\Phi}{n}{}{-1}{}\right)^2 \\-\frac{3}{8 \dd\mathrm{x}^2} \left(\lat{\Phi}{n}{}{}{+1}-\lat{\Phi}{n}{}{}{-1}\right)^2 - \frac{3\cH}{\dd\tau} \lat{\Phi}{n}{}{}{} -4 \pi G a^2 \left(1-4\lat{\Phi}{n}{}{}{}\right) \left[T^0_0(\taulat{+1}, \xlat{}{}{}) - \bar{T}^0_0(\taulat{+1})\right] \, .
\end{multline}
With this definition, the unknown field values of $\Phi$ at $\taulat{+1}$ are given implicitly by
\begin{equation}
\label{eq:discrete00b}
 \frac{\lat{\Phi}{n+1}{-1}{}{} + \lat{\Phi}{n+1}{+1}{}{} + \lat{\Phi}{n+1}{}{-1}{} + \lat{\Phi}{n+1}{}{+1}{} + \lat{\Phi}{n+1}{}{}{-1} + \lat{\Phi}{n+1}{}{}{+1} - 6 \lat{\Phi}{n+1}{}{}{}}{\dd\mathrm{x}^2} - \frac{3\cH}{\dd\tau} \lat{\Phi}{n+1}{}{}{} = \lat{s}{}{}{}{} \, .
\end{equation}
Note that this equation is linear in the unknown field values, and we can therefore invert the finite-difference operator
on the left hand side using Fourier analysis\footnote{Numerically this may not be the most efficient approach. For instance, the so-called multigrid algorithm is known to be superior in many
cases and can even be generalized in order to invert nonlinear operators. In our previous work \cite{Adamek:2014xba} we considered yet a different algorithm based on an operator splitting method
that performs an approximate inversion. It turned out that, once the symmetry assumption used in that work was dropped, the errors introduced by this approach were intolerable in practice.
We decided to use Fourier analysis (which is an exact method) because we use a similar solution strategy for eq.~(\ref{eq:ij}). The latter requires more execution time than eq.~(\ref{eq:00})
in any case, and optimizing on eq.~(\ref{eq:00}) would therefore not have a large impact on total performance.}.
The discrete Fourier transform
\begin{equation}
 \latFT{s}{} \doteq \sum_{\mathbf{i},\mathbf{j},\mathbf{k}} \lat{s}{}{}{}{} e^{-2\pi i \left(\mathbf{i}\mathbf{u} + \mathbf{j}\mathbf{v} + \mathbf{k}\mathbf{w}\right) / N}
\end{equation}
is computed using a distributed three dimensional FFT algorithm provided by the \lf\ library. Here, $\mathbf{u}$, $\mathbf{v}$, $\mathbf{w}$ are discrete indices labelling
the Fourier wave vector and $N$ denotes the number of lattice points in each dimension, the total lattice having $N^3$ points. In discrete Fourier space eq.~(\ref{eq:discrete00b}) becomes
\begin{equation}
 -\left(\frac{4}{\dd\mathrm{x}^2} \sin^2 \frac{\pi \mathbf{u}}{N} + \frac{4}{\dd\mathrm{x}^2} \sin^2 \frac{\pi \mathbf{v}}{N} + \frac{4}{\dd\mathrm{x}^2} \sin^2 \frac{\pi \mathbf{w}}{N} + \frac{3\cH}{\dd\tau} \right)\latFT{\Phi}{n+1} = \latFT{s}{} \, .
\end{equation}
As usual, the linear finite-difference operator in position space corresponds to a multiplicative function which depends on the Fourier wave vector. We can now simply
solve this algebraic equation for $\latFT{\Phi}{n+1}$ and perform the inverse FFT in order to obtain $\lat{\Phi}{n+1}{}{}{}$. At this point it is important to note that
the equation gives a meaningful value also for $\latFT{\Phi}{n+1}$ at $\mathbf{u}=\mathbf{v}=\mathbf{w}=0$. In other words, if $\lat{s}{}{}{}{}$ has a homogeneous mode it 
sources a homogeneous mode in $\Phi$. This property is crucial for internal consistency of the scheme. In the Newtonian limit, $\lat{s}{}{}{}{}$ does not have
a homogeneous mode by construction, and it is hence consistent to set the homogeneous mode of $\Phi$ to zero.

Next, we have to solve for the remaining perturbation variables $\chi$, $B_i$ and $h_{ij}$; we follow a very similar philosophy. We construct
a nonperturbative source tensor $S_{ij}$ by moving the nonlinear terms of eq.~(\ref{eq:ij}) to the right hand side. As explained in Appendix~\ref{app:PM}, scalar quantities like $\chi$ and $\Phi$
are represented at the vertices of the lattice, as are the diagonal components of spatial tensors, e.g.\ $h_{ii}$. The components of $B_i$ are represented
on the edges, and the off-diagonal elements of a spatial tensor on the faces of the lattice, respectively. This needs to be taken into account when constructing the
source tensor $S_{ij}$. In order to avoid too many messy equations, we give the construction of one diagonal element, $S_{11}(\xlat{}{}{})$, and one off-diagonal element, $S_{12}(\xlat{+\frac{1}{2}}{+\frac{1}{2}}{})$. The other components of the tensor are constructed analogously.
\begin{multline}
\label{eq:S11}
 S_{11}(\xlat{}{}{}) \doteq 8 \pi G a^2 T^1_1(\taulat{+1},\xlat{}{}{})\\
 - \frac{\left(\lat{\Phi}{n+1}{+1}{}{} - \lat{\Phi}{n+1}{-1}{}{}\right)^2}{2\dd\mathrm{x}^2} +\left(2 \lat{\chi}{n}{}{}{} - 4 \lat{\Phi}{n+1}{}{}{}\right) \frac{\lat{\Phi}{n+1}{+1}{}{} + \lat{\Phi}{n+1}{-1}{}{} - 2 \lat{\Phi}{n+1}{}{}{}}{\dd\mathrm{x}^2}  \, ,
\end{multline}
\begin{multline}
\label{eq:S12}
 S_{12}(\xlat{+\frac{1}{2}}{+\frac{1}{2}}{}) \doteq 8 \pi G a^2 T^1_2(\taulat{+1},\xlat{+\frac{1}{2}}{+\frac{1}{2}}{}) \\
 -\frac{1}{2} \, \frac{\lat{\Phi}{n+1}{+1}{+1}{} - \lat{\Phi}{n+1}{}{+1}{} + \lat{\Phi}{n+1}{+1}{}{} - \lat{\Phi}{n+1}{}{}{}}{\dd\mathrm{x}} \, \frac{\lat{\Phi}{n+1}{+1}{+1}{} - \lat{\Phi}{n+1}{+1}{}{} + \lat{\Phi}{n+1}{}{+1}{} - \lat{\Phi}{n+1}{}{}{}}{\dd\mathrm{x}} \\
 +\frac{1}{2} \left(\lat{\chi}{n}{}{}{} + \lat{\chi}{n}{+1}{}{} + \lat{\chi}{n}{}{+1}{} + \lat{\chi}{n}{+1}{+1}{}\right)\frac{\lat{\Phi}{n+1}{+1}{+1}{} - \lat{\Phi}{n+1}{}{+1}{} - \lat{\Phi}{n+1}{+1}{}{} + \lat{\Phi}{n+1}{}{}{}}{\dd\mathrm{x}^2} \\
 - \left(\lat{\Phi}{n+1}{}{}{} + \lat{\Phi}{n+1}{+1}{}{} + \lat{\Phi}{n+1}{}{+1}{} + \lat{\Phi}{n+1}{+1}{+1}{}\right)\frac{\lat{\Phi}{n+1}{+1}{+1}{} - \lat{\Phi}{n+1}{}{+1}{} - \lat{\Phi}{n+1}{+1}{}{} + \lat{\Phi}{n+1}{}{}{}}{\dd\mathrm{x}^2}  \, .
\end{multline}
We have written eq.~(\ref{eq:S12}) in a suggestive way such that the terms can be easily identified. For instance, the second line is the product $2 \Phi_{,1} \Phi_{,2}$
with the two gradients interpolated to the face of the lattice cell at $\xlat{+\frac{1}{2}}{+\frac{1}{2}}{}$. Within the code though, these expressions are rearranged in order to
minimize the number of operations required to construct $S_{ij}$. Note also that the terms proportional to $\chi$ in eqs.~(\ref{eq:S11})
and (\ref{eq:S12}) are neglected in the public version of the code. This is done for efficiency reasons and is justified as long as $\chi \ll \Phi$, since $\chi$ only appears in the combination $4\Phi-2\chi$ within these equations.

With $S_{ij}$, the continuum version of eq.~(\ref{eq:ij}) reads
\begin{equation}
 \frac{1}{2} h_{ij}'' + \cH h_{ij}' - \frac{1}{2} \Delta h_{ij} + B_{(i,j)}' + 2 \cH B_{(i,j)} + \chi_{,ij} - \frac{1}{3} \delta_{ij} \Delta \chi = S_{ij} - \frac{1}{3} \delta_{ij} \delta^{kl} S_{kl} \, ,
\end{equation}
which is linear in the unknown perturbation variables\footnote{The dependence of $S_{ij}$ on these variables is very subleading and for the purpose of solving this
equation we simply estimate their contributions using the values from the previous time step. This is fine within a first-order in time integration
scheme as the one we use here. If one were to move to a higher-order scheme, one could again employ a predictor-corrector method.}
$\chi$, $B_i$ and $h_{ij}$, meaning that we can again use Fourier analysis as a solution strategy. There is a great advantage to this approach: in Fourier
space, the separation into spin-0, spin-1, and spin-2 components is straightforward.

For the spin-0 component, we simply contract the indices with two Fourier wave vectors. The gauge conditions then imply that
\begin{equation}
\label{eq:chi}
 \tilde{\chi}(\mathbf{k}) = \frac{1}{2k^4} \left(\delta^{ij} k^2 - 3 k^i k^j\right) \tilde{S}_{ij}(\mathbf{k}) \, .
\end{equation}
The spin-1 component is projected out by applying the transverse projector $P^{ij} \doteq \delta^{ij} - k^i k^j / k^2$ on one of the indices and contracting the other one
again with a Fourier wave vector. We obtain
\begin{equation}
\label{eq:Bi}
 \tilde{B}_i'(\mathbf{k}) + 2 \cH \tilde{B}_i (\mathbf{k}) = \frac{1}{a^2}\left(a^2 \tilde{B}_i(\mathbf{k})\right)' = -\frac{2 i}{k^4} \delta_{ij} \left(\delta^{jl} k^2 - k^j k^l\right) k^m \tilde{S}_{lm}(\mathbf{k}) \, .
\end{equation}
Finally, for the spin-2 projection, we apply $\Lambda^{ijlm} \doteq P^{il} P^{jm} - \frac{1}{2} P^{ij} P^{lm}$, which yields
\begin{multline}
\label{eq:hij}
 \left(\tilde{h}_{lm}''(\mathbf{k}) + 2 \cH \tilde{h}_{lm}'(\mathbf{k}) + k^2 \tilde{h}_{lm}(\mathbf{k})\right) \delta^{il} \delta^{jm} = \\ \frac{2}{k^4} \left[\left( \delta^{il} k^2 - k^i k^l\right) \left(\delta^{jm} k^2 - k^j k^m\right) - \frac{1}{2} \left(\delta^{ij} k^2 - k^i k^j\right) \left(\delta^{lm} k^2 - k^l k^m\right)\right] \tilde{S}_{lm}(\mathbf{k}) \, .
\end{multline}
The equations (\ref{eq:chi}), (\ref{eq:Bi}) and (\ref{eq:hij}) are valid in the continuum limit where the Fourier wave vectors correspond to spatial derivatives in position space.
On the lattice, we have to replace these wave vectors by the algebraic expressions corresponding to the specific notion of finite-difference operators
chosen in the discretization procedure. For instance, one can find a version of $P^{ij}$ acting in discrete Fourier space in such a way that, after
inverse Fourier transform, the discrete divergence of $B_i$ vanishes exactly, satisfying the discrete version of the gauge condition.

Explicitly, we define the discrete Fourier components of $S_{ij}$ as
\begin{align}
 \tilde{S}_{11}(\mathbf{u}, \mathbf{v}, \mathbf{w}) \doteq &\, \sum_{\mathbf{i},\mathbf{j},\mathbf{k}} S_{11}(\xlat{}{}{}) e^{-2\pi i \left(\mathbf{i}\mathbf{u} + \mathbf{j}\mathbf{v} + \mathbf{k}\mathbf{w}\right) / N} \, , \\
 \tilde{S}_{12}(\mathbf{u}, \mathbf{v}, \mathbf{w}) \doteq &\, e^{-\pi i \mathbf{u}/N} e^{-\pi i \mathbf{v}/N}\sum_{\mathbf{i},\mathbf{j},\mathbf{k}} S_{11}(\xlat{+\frac{1}{2}}{+\frac{1}{2}}{}) e^{-2\pi i \left(\mathbf{i}\mathbf{u} + \mathbf{j}\mathbf{v} + \mathbf{k}\mathbf{w}\right) / N} \, ,
\end{align}
and similarly for the remaining components. Note that the off-diagonal elements are defined with a phase factor which accounts for the fact that these components
exist on the faces of the lattice cells and not on the vertices. Let us also define a lattice momentum $\klat{i}$ as
\begin{equation}
 \left(\klat{1}, \klat{2}, \klat{3}\right) \doteq \left(\frac{2}{\dd\mathrm{x}} \sin\frac{\pi \mathbf{u}}{N}, \frac{2}{\dd\mathrm{x}} \sin\frac{\pi \mathbf{v}}{N}, \frac{2}{\dd\mathrm{x}} \sin\frac{\pi \mathbf{w}}{N}\right) \, .
\end{equation}
With these definitions, the discrete versions of eqs.~(\ref{eq:chi}) -- (\ref{eq:hij}) are simply given by replacing $\tilde{S}_{ij}(\mathbf{k}) \rightarrow \tilde{S}_{ij}(\mathbf{u}, \mathbf{v}, \mathbf{w})$, $k^i \rightarrow \klat{i}$, $k^2 \rightarrow \delta_{ij} \klat{i} \klat{j}$, and so forth. The discrete Fourier transform of $h_{ij}$ is defined in the same way as the one for $S_{ij}$,
while the one for $B_i$ takes the form
\begin{equation}
 \tilde{B}_{1}(\mathbf{u}, \mathbf{v}, \mathbf{w}) \doteq e^{-\pi i \mathbf{u}/N} \sum_{\mathbf{i},\mathbf{j},\mathbf{k}} B_{1}(\xlat{+\frac{1}{2}}{}{}) e^{-2\pi i \left(\mathbf{i}\mathbf{u} + \mathbf{j}\mathbf{v} + \mathbf{k}\mathbf{w}\right) / N} \, ,
\end{equation}
and similarly for the other components, taking into account the fact that the components of a vector reside on the edges of the lattice cells.

In order to evolve $B_i$, the code keeps a copy of $\mathcal{B}_i \doteq a^2 B_i$ in Fourier space. According to eq.~(\ref{eq:Bi}), the update step for $\mathcal{B}_i$ is then
performed as
\begin{equation}
 \tilde{\mathcal{B}}_i^{\mathbf{n+1}}(\mathbf{u}, \mathbf{v}, \mathbf{w}) = \tilde{\mathcal{B}}_i^{\mathbf{n}}(\mathbf{u}, \mathbf{v}, \mathbf{w}) - \frac{2 i a^2 \dd\tau}{\klat{4}} \delta_{ij} \left(\delta^{jl} \klat{2} - \klat{j} \klat{l}\right) \klat{m} \tilde{S}_{lm}(\mathbf{u}, \mathbf{v}, \mathbf{w}) \, .
\end{equation}

Even though we have all the ingredients ready for solving eq.~(\ref{eq:hij}), in the present version of the code this is not what we do.
The reason is that, for the scenarios studied in this work,
the scattering of particles with the spin-2 metric perturbation $h_{ij}$ is a very small effect and is therefore neglected.
The effect is much smaller than the frame dragging caused by $B_i$,
which is taken into account in the code and which is already much smaller than the gravitational forces due to the potentials.
For the purpose of evolution we therefore ignore the backreaction from
$h_{ij}$. This is not a limitation of our framework as one could easily take it into account\footnote{Actually, in $\Lambda$CDM the amplitude
of $h_{ij}$ coming from large scale structure is so tiny that its effect on the geodesic equation or the geometric factors in the
particle-to-mesh projection would be drowned in the numerical rounding errors when working at single precision (four-byte floating point numbers).}.

In the case where $h_{ij}$ is (mainly) sourced through non-relativistic matter, we can still reconstruct it at any time using the quasi-static
approximation. This approximation relies on the fact
that the sources (and the potentials) evolve slowly such that the solution of eq.~(\ref{eq:hij}) adiabatically follows the equilibrium
configuration, at least at scales well inside the horizon.
The equilibrium configuration is defined through eq.~(\ref{eq:hij}) by neglecting the time derivatives of $h_{ij}$. It should be
noted that $h_{ij}$ does not describe a free gravitational wave
in this case. It corresponds to a static spin-2 perturbation of the geometry. Oscillations on top of this equilibrium
configuration only occur through violation of the adiabatic approximation.

For $\Lambda$CDM the quasi-static approximation for the gravitational wave spectrum is certainly sufficient. To go beyond it is
however not trivial because contrary to $B_i$, $\Phi$ and $\chi$, the non-quasistatic contribution to $h_{ij}$ varies rapidly in time. One possibility is to update the rapidly oscillating (but very small) tensor fluctuations in much shorter time steps
than all the other variables. Another way would be to store the unequal time correlator of the tensor anisotropic stress in Fourier
space, $\Pi^{T}(k,\tau,\tau')$, and  compute the gravitational wave spectrum with the Wronskian method,
\begin{equation}
P_h(k,\tau_0) =\int_{\tau_{\rm in}}^{\tau_0}d\tau \int_{\tau_{\rm in}}^{\tau_0}d\tau' G(k,\tau_0,\tau)G(k,\tau_0,\tau')\Pi^{T}(k,\tau,\tau') \,,
\end{equation}
where $G(k,\tau_0,\tau')$ is the Green function for the tensor differential operator.
This method has not been implemented in the present version of the code as it would require to compute the source unequal time correlator at many time steps
which is relatively expensive.

\bibliographystyle{JHEP}
\bibliography{gevolution}

\end{document}